\documentclass[journal,10pt]{IEEEtran}
\usepackage{amsmath}
\usepackage{amsfonts}
\usepackage{amssymb}
\usepackage{algorithm}
\usepackage{algorithmic}
\usepackage{paralist}
\usepackage{graphicx}
\usepackage{booktabs}
\usepackage{multirow}
\usepackage{array}
\usepackage{stfloats}
\usepackage{subfig}
\usepackage{cite}
\usepackage{bbm}

\usepackage{lipsum} 
\usepackage{amsmath,epsfig,amssymb,bm,dsfont}
\allowdisplaybreaks[0] 
\usepackage{setspace}
\ifCLASSINFOpdf
\else
\fi
\allowdisplaybreaks[4]

\begin{document}

\newtheorem{theorem}{Theorem}
\newtheorem{proposition}{Proposition}
\newtheorem{lemma}{Lemma}
\newtheorem{definition}{Definition}
\newtheorem{proof}{Proof}

\title{Joint Sparse Beamforming and Power Control for a Large-scale DAS with Network-Assisted Full Duplex}
\author{Xinjiang~Xia$^{1}$,~\IEEEmembership{}
        Pengcheng~Zhu$^{1}$,~\IEEEmembership{Member,~IEEE,}
        Jiamin~Li$^{1}$,~\IEEEmembership{}
        Dongming~Wang$^{1}$,~\IEEEmembership{Member,~IEEE,}
        Yuanxue~Xin$^{2}$,~\IEEEmembership{}
        and~Xiaohu~You$^{1}$,~\IEEEmembership{Fellow,~IEEE}\\
         $^{1}$National Mobile Communications Research Laboratory, Southeast University, China\\
$^{2}$College of Internet of Things Engineering, Hohai University, China\\
\thanks{This work was supported in part by the National Natural Science Foundation of China (NSFC) under Grant 61871122, Grant 61801168,
and Grant 61571120, in part by the National Key Special Program under Grant 2018ZX03001008-002, and in part by the Six Talent Peaks
Project in Jiangsu Province. }
}
\maketitle
\begin{abstract}
In this paper, we studied joint sparse beamforming and power control for network-assisted full duplex (NAFD) in a large-scale distributed antenna system (L-DAS), where the remote antenna units are operating in either half-duplex mode or full-duplex mode and are all connected to the central processing unit (CPU) via high-speed backhaul links. With joint processing at the CPU, NAFD can achieve truly flexible duplex, including flexible half-duplex and full-duplex modes. Cross-link interference and finite-capacity backhaul are the main problems of NAFD in an L-DAS. To solve these problems, we aim to maximize the aggregated uplink and downlink rates subject to quality-of-service constraints and backhaul constraints. Two approaches have been proposed to solve the optimization problem, where the first approach converts the objective function to the difference of two convex functions with semi-definite relaxation (SDR), and then, an iterative SDR-block coordinate descent method is applied to solve the problem.
The second method is based on sequential parametric convex approximation. Simulation results have shown that the proposed algorithms yield a higher rate gain compared to the traditional time-division duplex scheme.
\end{abstract}

\begin{IEEEkeywords}
sparse beamforming,  power control, network-assisted full duplex, Large-scale DAS (L-DAS), semi-definite relaxation-block coordinate descent (SDR-BCD), sequential parametric convex approximation (SPCA).
\end{IEEEkeywords}

\IEEEpeerreviewmaketitle

\section{Introduction}

Over the last decade, the proliferation of smart devices and video streaming applications has led to an explosive rise in the demand for a higher data rate of both the uplink and downlink in cellular systems \cite{sun2015d2d}. To improve the spectral efficiency and reduce latency, the fifth generation new radio supports flexible duplex technology in both the paired and unpaired spectrum, i.e., it allows  frequency division duplex on a paired spectrum and time division duplex (TDD) operation on an unpaired spectrum. With co-frequency co-time full duplex (CCFD), a wireless transceiver can simultaneously transmit and receive in the same frequency band, and thus, the spectral efficiency could be further doubled by effectively eliminating self-interference \cite{sabharwal2014band,nguyen2014spectral,li2017spectral,bi2018fractional,song2017antenna}. Therefore, CCFD is one of the techniques enabling the sixth generation.

In addition to CCFD, researchers have also studied the spatial-domain duplex techniques. In \cite{thomsen2016compflex}, a spatial domain duplex was suggested in which one user terminal receives the downlink from a half-duplex base station (BS), and the second user terminal transmits to another half-duplex BS on the same time-frequency resource blocks. In \cite{xin2017antenna}, a spatial domain duplexing technique called the bidirectional dynamic network (BDN) was proposed for large-scale distributed antenna systems (L-DASs). In BDN, uplink users (UUs) and downlink users (DUs) are associated with different remote antenna units (RAUs). By dynamically allocating the number of transmitting RAUs (T-RAUs) and receiving RAUs (R-RAUs), both uplink and downlink service could be supported simultaneously and flexibly.

Cross-link interference (CLI), that is, the interference from uplink users to downlink users and the interference from transmitting BSs to receiving BSs, is the most difficult problem for cellular networks with CCFD, flexible duplex or BDN. In \cite{Dongming2019}, network-assisted full duplex (NAFD) based on an L-DAS was proposed to reduce CLI by using joint processing. NAFD could be viewed as a unified implementation of flexible duplex, CCFD and hybrid duplex under a cell-free network architecture, where all the RAUs are connected to the central processing unit (CPU) via high-speed backhaul.

For an L-DAS with NAFD, there are two problems that make its application very challenging. First, the downlink-to-uplink interference relies on the joint processing, which requires a large amount of backhaul. Second, the CLI should be controlled to be very low; otherwise, it will also have a negative effect on the quality-of-service (QoS) of both uplink users and downlink users.

In recent years, cellular systems with finite-capacity backhaul links have been studied by many researchers \cite{dai2014sparse,dai2016energy,tabassum2016analysis}. In \cite{dai2014sparse}, network utility maximization for the downlink cloud radio access network (C-RAN) with per-RAU backhaul capacity constraints was investigated, and network energy efficiency maximization for downlink transmission with load-dependent backhaul power was considered in \cite{dai2016energy}. The asymptotic performance of massive MIMO-enabled wireless backhaul nodes with  CCFD small cells operating in either the in-band or out-of-band mode was analysed in \cite{tabassum2016analysis}.

The CLI in NAFD will make some users have a lower spectral efficiency. In \cite{dong2015energy,cirik2018fronthaul,li2018spectral}, the researchers proposed some optimization algorithms for CCFD to guarantee QoS of both downlink users and uplink users. Since the uplink receiver and the downlink beamforming are coupled in the optimization problems, the transceiver design with QoS constraints in CCFD systems is more complex than that in half-duplex systems\cite{chen2016green,jiang2017max,tan2018virtual}. Although with joint processing, the inter-RAU interference (IRI) could be suppressed, the residual IRI due to imperfect cancellation should be considered.
Furthermore, in addition to IRI, there still exists inter-user interference among uplink users and downlink users, and therefore, uplink power control should be investigated.

In this paper, we studied the joint downlink beamforming, uplink receiver, and uplink power control for NAFD in an L-DAS with finite-capacity backhaul and QoS constraints. The contributions of this paper are summarized as follows:
\begin{itemize}
\item We developed a general optimization framework of transceiver design for NAFD with imperfect downlink-to-uplink interference cancellation. Using the semi-definite relaxation block coordinate descent (SDR-BCD) algorithm, we could solve the maximization problem of the total uplink and downlink rate under power, QoS and backhaul constraints. To the best of the authors�� knowledge, this work is the first attempt to study the joint sparse beamforming and power control with backhaul constraints for NAFD.
\item We also proposed a two-stage algorithm to solve the sparse optimization problem. In the first stage, we provided a user selection algorithm by approximating the indicator function of DU-RAU association with a continuous function. Then, in the second stage, we presented a joint beamforming and power control algorithm.
\item To further reduce the complexity, we proposed an iterative algorithm based on sequential parametric convex approximation (SPCA).
\end{itemize}

The remainder of this paper is organized as follows. Section II presents the channel model, problem formulation and
system model. Then, we propose the SDR-BCD-based algorithm for an L-DAS with NAFD in Section III.
In Section IV, we jointly optimize the transceiver parameters based on SPCA.
The performance of the proposed algorithms is evaluated by simulations in Section V.
Finally, Section VI summarizes the paper.

Notations: Throughout this paper, scalars are represented by lower-case letters (e.g., $i$), matrices are represented by upper-case bold letters (e.g., $\bm{H}$), and vectors  are represented by lower-case bold letters (e.g., $\bm{v}$). The matrix inverse, conjugate transpose and ${l_p}$-norm of a vector are denoted ${\left( . \right)^{ - 1}}$, ${\left( . \right)^{\rm H}}$ and ${\left| . \right|_p}$ respectively.  ${\mathbb{C}^{M \times N}}$ is used to denote the set of complex $M \times N$ matrices. The complex Gaussian distribution is represented by $ \mathcal{CN}(.,.)$. Calligraphy letters are used to denote sets.

\begin{figure}[!htb]
\begin{center}
\includegraphics[scale=0.6]{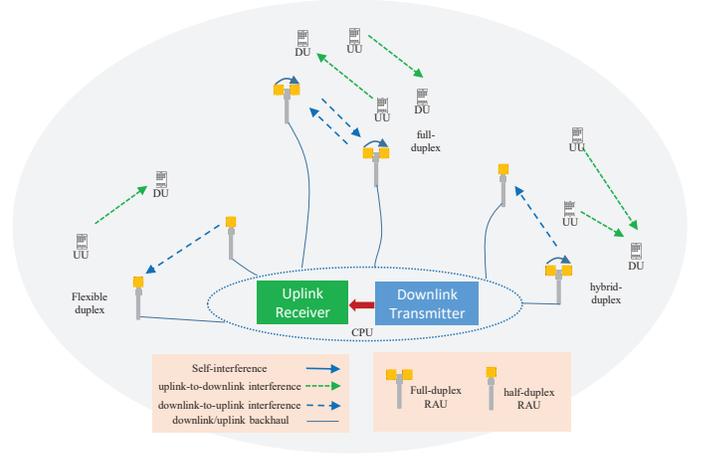}
\caption{Illustration of an L-DAS with NAFD.}\label{FDN}
\end{center}
\end{figure}

\section{System Model and Problem Formulation}

\subsection{Network-Assisted Full Duplex for an L-DAS}
The system model of an L-DAS with NAFD is illustrated in Fig. \ref{FDN}. The RAUs can be either CCFD or half-duplex. For a CCFD RAU, theoretically, it could be considered as two RAUs: one is for uplink reception (R-RAU) and the other for downlink transmission (T-RAU). It is assumed that for a CCFD RAU, the self-interference could be mostly cancelled in the analog domain, and the residual self-interference will be modelled in the following. It is also assumed that the user terminals are half-duplex because of the limited hardware capability, and both uplink users and downlink users are working on the same time-frequency resources.

In an L-DAS with NAFD, the CPU generates multi-user beamforming signals of downlink users and sends them to T-RAUs via downlink backhaul, and on the same time-frequency resources, the CPU also receives uplink signals of uplink users from R-RAUs via uplink backhaul and then performs joint multi-user detection. Furthermore, we assume that the uplink users' data can be detected by only one R-RAU; thus, compared with downlink, uplink usually has less traffic.
If the downlink backhaul is satisfied, then the uplink transmission will also be satisfied.

\subsection{Signal Model}
We consider an NAFD system that contains $L$ T-RAUs, $Z$ R-RAUs, $K$ downlink users and $J$ uplink users. Each RAU has $M$ antennas, and each user has only one antenna. Let $\mathcal{L} = \left\{ {1, \cdots ,L} \right\}$ and $\mathcal{K} = \left\{ {1, \cdots ,K} \right\}$ denote the sets of T-RAU and downlink user indices, respectively. Let $\mathcal{Z} = \left\{ {1, \cdots ,Z} \right\}$ and
$\mathcal{J} = \left\{ {1, \cdots ,J} \right\}$ denote the sets of R-RAU and uplink user indices, respectively.

Let us first consider the downlink. We model the received signal of downlink user $k$ as
\begin{equation}\label{signD}
\begin{aligned}
y_{\text{D},k} = \sum\limits_{k' \in \mathcal{K}} { {\bm{h}}^{\rm H}_{\text{D},k} {\bm{w}}_{\text{D},k'}s_{\text{D},k'}} + \sum\limits_{j \in \mathcal{J}} {{h}_{\text{IUI},{j,k}}\sqrt {P_{\text{U},j}} s_{\text{U},j} + n_{\text{D},k}},
\end{aligned}
\end{equation}
where
${\bm{w}}_{\text{D},k} = {\left[ {{{ {{\bm{w}}^{\rm T}_{\text{D},{1,k}}} }}, \ldots ,{{ {{\bm{w}}^{\rm T}_{\text{D},{L,k}}} }}} \right]^{\rm T}}{\rm{}} \in\mathbb{C} {^{ML \times 1}}$
denotes the beamforming vector for the data streams of downlink user $k$,
${\bm{h}}_{\text{D},k}= {\left[ {{{ {{\bm{h}}^{\rm T}_{\text{D},{1,k}}} }},{{ {{\bm{h}}^{\rm T}_{\text{D},{2,k}}} }} \cdots {{ {{\bm{h}}^{\rm T}_{\text{D},{L,k}}} }}} \right]^{\rm T}} \in\mathbb{C} {^{ML \times 1}}$
denotes the channel vector from all T-RAUs to downlink user $k$, ${{s}}_{\text{D},k}\sim {\cal C}{\cal N}(0, 1)$ is the intended signal for downlink user $k$, ${{s}}_{\text{U},j}\sim {\cal C}{\cal N}(0, 1)$ is the signal of uplink user $j$, $ n_{\text{D},k} \sim {\cal C}{\cal N}(0,\sigma _{{\rm{D}},k}^2)$ is the additive white Gaussian noise, ${{h}}_{\text{IUI},{j,k}}$ denotes the channel coefficient of inter-user-interference (IUI) from uplink user $j$ to downlink user $k$, and $P_{\text{U},j}$ is the uplink transmission power of uplink user $j$. Then, we can write the rate of downlink user $k$ as
\begin{equation}\label{rateD}
\begin{aligned}
R_{\text{D},k} = {\log _2}\left( {1 + {r_{\text{D},k}}} \right) =\log_2 \bigg( {1 + \frac{{{{\left| { {\bm{h}}^{\rm H}_{\text{D},k} {\bm{w}}_{\text{D},k}} \right|}^2  }}}{{\gamma }_{\text{D},k}}} \bigg),
\end{aligned}
\end{equation}
where
\begin{equation}\label{rateDD}
\begin{aligned}
{{\gamma }}_{\text{D},k}={{\sum\limits_{{k \in \mathcal{K}}, k' \ne k} {{{\left| { {\bm{h}}^{\rm H}_{\text{D},k} {\bm{w}}_{\text{D},k'}} \right|}^2}}  + \sum\limits_{j \in \mathcal{J}} {P_{\text{U},j}{{\left| {{{h}}_{\text{IUI},{j,k}}} \right|}^2} + } \sigma _{\text{D},k}^2}} \nonumber
\end{aligned}
\end{equation}
is the variance in the interference plus noise at receiver $k$. We can see that to improve the downlink performance, we should control the transmission power of uplink users and thus reduce the interference from uplink users.

For the uplink, we first define $\tilde {\cal J}_z \in {\cal J}$, representing the set of user indices served by R-RAU $z$, and $\left| {\tilde {\cal J}_z} \right|$, representing the number of users served by R-RAU $z$. Furthermore, we assume that the uplink user and R-RAU pairs have been fixed. Then, we can write the received signal of R-RAU $z$ as
\begin{small}
\begin{equation}\label{signURAU}
\begin{aligned}
{\bm{y}}_{\text{U},z} = \sum\limits_{j \in \mathcal{J}} {{\bm{h}}_{\text{U},{j,z}}\sqrt {P_{\text{U},j}} s_{\text{U},j}}  + \sum\limits_{k \in \mathcal{K}} {{\bm{H}}_{\text{IRI},z}{\bm{w}}_{\text{D},k}s_{\text{D},k}}  + {\bm{n}}_{\text{U},z},
\end{aligned}
\end{equation}
\end{small}
where ${\bm{h}}_{\text{U},{j,z}} \in\mathbb{C} {^{M \times 1}}$ denotes the channel vector from uplink user $j$ to R-RAU $z$, ${\bm{H}}_{\text{IRI},z} \in\mathbb{C} {^{M \times ML}} $ denotes the channel matrix from all T-RAUs to R-RAU $z$, i.e., the IRI channel between all T-RAUs and R-RAU $z$, ${\bm{n}}_{\text{U},z}$ denotes the additive Gaussian noise with zero mean and covariance matrix ${ {\sigma^2_{\text{U},z}} }{{\bm{I}}_M}$, and $s_{\text{U},j}$ represents the signal transmitted by uplink user $j$.

Theoretically, if the CPU has perfect channel state information between R-RAUs and T-RAUs (all ${\bm{H}}_{\text{IRI},{l,z}}$), the downlink-to-uplink interference (the second term on the right-hand-side (RHS) of (\ref{signURAU})) could be cancelled. However, in practice, it is difficult to obtain accurate ${\bm{H}}_{\text{IRI},{l,z}}$. We model the inter-RAU channel ${\bm{H}}_{\text{IRI},{l,z}}$  as
${\bm{H}}_{\text{IRI},{l,z}} = {\bm{\hat H}}_{\text{IRI},{l,z}} +  {\bm{\tilde{H}}}_{\text{IRI},{l,z}}$,
where ${\bm{H}}_{\text{IRI},{l,z}}  \in\mathbb{C} {^{M \times M}}$ denotes the imperfect channel between T-RAU $l$ and R-RAU $z$, ${\bm{\hat H}}_{\text{IRI},{l,z}}$ denotes the corresponding estimated channel and ${\bm{\tilde{H}}}_{\text{IRI},{l,z}}$ denotes the channel estimation error. We further assume that
the elements of ${\bm{\tilde{H}}}_{\text{IRI},{l,z}}$ follow a Gaussian distribution, i.e.,
$\text{vec}({\bm{\tilde{H}}}_{\text{IRI},{l,z}})  \sim \mathcal{CN} \left( {{\bm{0, }} \delta^2 _{{\text{IRI}},{l,z}} {{\bm{I}}_{{M}^2}}} \right),$
where $ \delta^2 _{{\text{IRI}},{l,z}} $ denotes the residual interference power due to imperfect IRI cancellation in the digital or analog domain.

After proper interference cancellation, the signal received by  RAU $z$ can be modelled as
\begin{equation}\label{signUCP}
\begin{aligned}
{{{\bm{\hat y}}}_{{\rm{U}},z}} = \sum\limits_{j \in \mathcal{J}} {{{\bm{h}}_{{\rm{U}},j,z}}\sqrt {{P_{{\rm{U}},j}}} {s_{{\rm{U}},j}}}  + \sum\limits_{k \in \mathcal{K}} { {{\bm{\tilde{H}}}_{{\rm{IRI}},z}}{{\bm{w}}_{{\rm{D}},k}}{s_{{\rm{D}},k}}}  + {{\bm{n}}_{{\rm{U}},z}},
\end{aligned}
\end{equation}
where ${\bm{\tilde{H}}}_{\text{IRI},z} = {\left[ {{{ {{\bm{\tilde{H}}}^{\rm T}_{\text{IRI},{1,z}}} }},{{ {{\bm{\tilde{H}}}^{\rm T}_{\text{IRI},{2,z}}} }}, \cdots {{ {{\bm{\tilde{H}}}^{\rm T}_{\text{IRI},{l,z}}} }}} \right]^{\rm T}}$.

The covariance matrix of IRI between  R-RAU $z$ and T-RAU $l$  can be given by
\begin{equation}\label{signzlSI}
\begin{aligned}
\mathbb {E}\left[{{\bm{\tilde{H}}}_{\text{IRI},{l,z}}{\bm{w}}_{\text{D},{l,k}}{ {{\bm{w}}^{\rm H}_{\text{D},{l,k}}} }{ {{\bm{\tilde{H}}}^{\rm H}_{\text{IRI},{l,z}}} }} \right]
=  \delta^2 _{{\text{IRI}},{l,z}}  {{\left\| {{{\bm{w}}_{{\rm{D}},l,k}}} \right\|}^2}{{\bm{I}}_M}.
\end{aligned}
\end{equation}

To detect $s_{\text{U},j}$, we assume that RAU $z$ employs single-user detection after applying a receive beamforming vector ${{\bm{u}}_{\text{U},j,z}} \in\mathbb{C} {^{M \times 1}}$ to ${{{\bm{\hat y}}}_{\text{U},z}}$ for $j \in \tilde {{\cal J}_{z}}$, and ${{\bm{u}}_{{\rm{U}},z}} = \left\{ {{{\bm{u}}_{{\rm{U}},j,z}}} \right\}_{j = 1}^J\in\mathbb{C} {^{JM \times 1}}$; then, the uplink rate for each uplink user $j$ can be expressed as
\begin{small}
\begin{equation}\label{rateU}
\begin{aligned}
&R_{\text{U},j} = {\log _2}\left( {1 + {r_{\text{U},j}}} \right)
 = {\log _2}\Big( {1 + \frac{{{P_{{\rm{U}},j}}{{\left| {{\bm{u}}_{{\rm{U}},j,z}^{\rm H}{{\bm{h}}_{{\rm{U}},j,z}}} \right|}^2}}}{{{\gamma_{{\rm{U}},j}}}}} \Big),
\end{aligned}
\end{equation}
\end{small}
where
\begin{equation}\label{ratezU1}
\begin{aligned}
{\gamma _{{\rm{U}},j}} =& {\rm{ }}\sum\limits_{{j \in \mathcal{J}}, j' \ne j} {{P_{{\rm{U}},j'}}{{\left| {{\bm{u}}_{{\rm{U}},j,z}^{\rm H}{{\bm{h}}_{{\rm{U}},j',z}}} \right|}^2}}  + \sigma _{{\rm{U}},z}^2{\left\| {{{\bm{u}}_{{\rm{U}},j,z}}} \right\|^2} \\
&+ \sum\limits_{k \in \mathcal{K}} { {\sum\limits_{l \in \mathcal{L}} {{{\left\| {{{\bm{w}}_{{\rm{D}},l,k}}} \right\|}^2}} \delta _{{\rm{IRI}},l,z}^2} } {\left\| {{{\bm{u}}_{{\rm{U}},j,z}}} \right\|^2}
\end{aligned}
\end{equation}
is the power of interference-plus-noise for uplink user $j$.

\subsection{Problem Formulation}

We aim to jointly optimize $\left\{ {P_{\text{U},j},{\bm{u}}_{\text{U},j,z},{\bm{w}}_{\text{D},k}} \right\}$ to maximize the sum-rate of the system under power and QoS constraints. The design problem is given by
\begin{subequations}\label{MSR}
\begin{align}
{\mathop {\max }\limits_{{\bm{w}}_{\text{D},k},{\bm{u}}_{\text{U},j,z},{P}_{\text{U},j}} }&{\sum\limits_{k \in \mathcal{K}} {R_{\text{D},k}} + \sum\limits_{j \in \mathcal{J}}{R_{\text{U},j}} }\\ \label{MSR1}
\text{s.t. } \quad&\sum\limits_{k \in \mathcal{K}} {{{ {{\bm{w}}^{\rm H}_{\text{D},{l,k}}} }}{\bm{w}}_{\text{D},{l,k}}}  \le {{\bar P}}_{\text{D},l}, \forall l,\\ \label{MSR2}
& 0 \le P_{\text{U},j} \le {{\bar P}_{\text{U},j}},\;\forall j,\\ \label{MSR3}
&\sum\limits_{k \in \mathcal{K}} \mathbbm {1}{\left\{ {\left\| {{\bm{w}}_{\text{D},{l,k}}} \right\|_2^2} \right\}} R_{\text{D},k} \le C_{\text{D},l},\;\forall l,\\ \label{MSR4}
&{R_{\text{D},k}}  \ge  {R_{{\text{D},\text{min}},k}},\;\forall k,\\ \label{MSR5}
&{R_{\text{U},j}}  \ge  {R_{{\text{U},\text{min}},j}},\;\forall j,\\ \label{MSR6}
&{{\bm{u}}_{{\rm{U}},j',z}} = {\bm{0}},\;\forall  j' \notin \tilde {\cal J}_z,
\end{align}
\end{subequations}
where ${{\bar P}}_{\text{D},l}$, ${{\bar P}_{\text{U},j}}$ and ${C_{\text{D},l}}$ are the power consumption budgets for T-RAU $l$ and uplink user $j$ and the backhaul constraint for T-RAU $l$.
\eqref{MSR4} and \eqref{MSR5} are the QoS constraints for downlink user $k$ and uplink user $j$, respectively. Condition \eqref{MSR6} imposes that the receiver should be a
zero vector when the index $j'$ is not served by RAU $z$.
\eqref{MSR3} corresponds to the backhaul constraints, where $\mathbbm {1}{\left\{ {\left\| {{\bm{w}}_{\text{D},{l,k}}} \right\|_2^2} \right\}}$ represents the indicator function that, with the facility of scheduling choice, is given by
\begin{equation}\label{indicator}
\mathbbm {1}{\left\{ {\left\| {{\bm{w}}_{\text{D},{l,k}}} \right\|_2^2} \right\}}=\left\{\begin{array}{ll}
0,&\text{if}\text{ }  {\left\| {{\bm{w}}_{\text{D},{l,k}}} \right\|_2^2} = 0,\\
1,&\text{otherwise}.
\end{array}\right.\
\end{equation}

As we can see, \eqref{indicator} is a non-convex expression. Motivated by the method used in \cite{liu2017cross}, we use the continuous function to approximate:
${\mathchar'26\mkern-10mu\lambda   _\theta }\left( {{\bm{w}}_{\text{D},{l,k}}} \right) = 1 - {e^{ - \theta \left\| {{\bm{w}}_{\text{D},{l,k}}} \right\|_2^2}}$,
where $\theta  \gg 1$. 

Then, we can rewrite problem \eqref{MSR} as
\begin{subequations}\label{MSRAS1}
\begin{align}
{\mathop {\max }\limits_{{\bm{w}}_{\text{D},k},{\bm{u}}_{\text{U},j,z},{P}_{\text{U},j}} }\text{ }&{\sum\limits_{k \in \mathcal{K}} {R_{\text{D},k}} + \sum\limits_{j \in \mathcal{J}}{R_{\text{U},j}} } \\ \label{MSRAS11}
\text{s.t. } \quad& \sum\limits_{k \in \mathcal{K}} {\mathchar'26\mkern-10mu\lambda _\theta }\left( {{\bm{w}}_{\text{D},{l,k}}} \right)  R_{\text{D},k} \le C_{\text{D},l}\\
& {\rm and} \quad \eqref{MSR1}, \text{ }\eqref{MSR2}, \text{ }\eqref{MSR4}, \text{ }\eqref{MSR5}, \text{ }\eqref{MSR6}.\nonumber
\end{align}
\end{subequations}

It is obvious that ${\mathchar'26\mkern-10mu\lambda _\theta }\big( {{\bm{w}}_{\text{D},{l,k}}} \big)$ is strictly less than $1$ when $\left\| {{\bm{w}}_{\text{D},{l,k}}} \right\|_2^2>0$.
Therefore, the solution of \eqref{MSRAS1} does not always satisfy constraint \eqref{MSR3} in original problem \eqref{MSR}. Since we assume that QoS is the basic requirement and the proper user selection algorithm has been applied, the constraints \eqref{MSR4} and \eqref{MSR5} always satisfy both the approximate problem and the original problem here.
This observation motivates us to solve problem \eqref{MSR} with two steps.

In the first step, we solve problem \eqref{MSRAS1} to obtain the beamforming vectors ${{\bm{\bar{w}}}_{\text{D},{l,k}}}$ and obtain the user-RAU association as
\begin{equation}\label{backhaul}
\begin{aligned}
\psi \left( {{\bm{\bar w}}_{\text{D},{l,k}}} \right) = \left\{ \begin{array}{ll}
1,&\text{if}\text{ }  {\mathchar'26\mkern-10mu\lambda _\theta }\left( {{\bm{\bar w}}_{\text{D},{l,k}}} \right) > \xi ,\\
0,&\text{otherwise}.
\end{array} \right.
\end{aligned}
\end{equation}
where $0 \le \xi \le 1$  denotes a threshold to control the user association.

In the second step, we solve the following problem to obtain the sparse beamforming and uplink power control:
\begin{subequations}\label{MSRAS2}
\begin{align}
& {\mathop {\max }\limits_{{\bm{w}}_{\text{D},k},{\bm{u}}_{\text{U},j,z},{P}_{\text{U},j}} }\text{ }{\sum\limits_{k \in \mathcal{K}} {R_{\text{D},k}} + \sum\limits_{j \in \mathcal{J}}{R_{\text{U},j}} }\\ \label{MSRAS21}
\text{s.t. } \quad&  \sum\limits_{k \in \mathcal{K}} \psi \left( {{\bm{\bar w}}_{\text{D},{l,k}}} \right)  R_{\text{D},k} \le C_{\text{D},l}, \forall l, \forall k,\\ \label{MSRAS22}
&{\left\| {{\bm{w}}_{\text{D},{l,k}}} \right\|_2^2}=0, \quad \forall \psi \left( {{\bm{\bar w}}_{\text{D},{l,k}}} \right)=0, \forall l, \forall k \\ \label{MSRAS23}
& {\rm and}\quad\eqref{MSR1}, \text{ }\eqref{MSR2}, \text{ }\eqref{MSR4}, \text{ }\eqref{MSR5}, \text{ }\eqref{MSR6}.\nonumber
\end{align}
\end{subequations}

Since the optimized parameters, such as downlink sparse beamformers, the uplink receiver, and the uplink transmission power, are tightly coupled in both subject function and constraints, the downlink and uplink transmission strategies should be optimized jointly. Note that problem \eqref{MSR} is a non-convex problem and difficult to solve. In the following, we try to address problem \eqref{MSR} with different approaches.

\section{Iterative SDR-BCD-based Algorithm}
To solve the coupled problem, we propose an iterative SDR-BCD-based algorithm. The main idea of the proposed algorithm is that when $\bm{u}_\text{U}$ is fixed, most of the expression of problem \eqref{MSR} can be solved by using a linear approximation, and when $\left\{ {P_{\text{U},j}, {\bm{w}}_{\text{D},k}} \right\}$ are fixed, $\bm{u}_\text{U}$  can be iteratively updated with a simple closed-form expression.
 
 Based on the description of Section \uppercase\expandafter{\romannumeral2}, we formulate the iterative SDR-BCD algorithm into a two-stage optimization problem to solve problems \eqref{MSRAS1} and \eqref{MSRAS2}.

\subsection{Stage \uppercase\expandafter{\romannumeral1}: Solution to Problem \eqref{MSRAS1} with the Iterative SDR-BCD-based Algorithm}
First, we define $\bm{Q}_{\text{D},k} =  {{\bm{w}}_{\text{D},k}} {{{\bm{w}}^{\rm H}_{\text{D},k}}}$ and define a set of matrices $\left\{ {{\bm{T}_l}} \right\}_{l = 1}^L$, where
each $\left\{ {{\bm{T}_l}} \right\} \in \left\{ 0,1  \right\} {^{M \times ML}}$ has the  form:
${{\bm{T}}_l} = \text{diag}\left[ { {{\bm{0}}_{M(l-1)}},  {{{\bm{1}}_M}},   {{\bm{0}}_{M(L-l)}} } \right]$.
Then, we can rewrite $\left\| {{\bm{w}}_{\text{D},{l,k}}} \right\|_2^2$ as
$\left\| {{\bm{w}}_{\text{D},{l,k}}} \right\|_2^2 = \text{Tr}\left( {{\bm{Q}}_{\text{D},k}{{\bm{T}}_l}} \right)$.
By removing the rank constraint, $\text{rank} \big( \bm{Q}_{\text{D},k}\big)=1$, we introduce the proposed algorithm as follows. With fixed ${\bm{u}}_{\text{U},j,z}$, \eqref{MSR} can be rewritten as
\begin{subequations}\label{MSRQ}
\begin{align}
&{\mathop {\max }\limits_{{\bm{Q}}_{\text{D},k},{P}_{\text{U},j}} }{\sum\limits_{k \in \mathcal{K}} {{\tilde {R}}_{\text{D},k}} + \sum\limits_{j \in \mathcal{J}}{{\tilde {R}}_{\text{U},j}} }\\  \label{MSRQ1}
&\text{s.t.}\;\;\sum\limits_{k \in \mathcal{K}} {{\rm{Tr}}\left( {{\bm{Q}}_{\text{D},k}{{\bm{T}}_l}} \right)}  \le {{\bar P}}_{\text{D},l}{\rm{ }}, \forall l, \forall k,\\ \label{MSRQ2}
&\;\;\;\; \;\;{\bm{Q}}_{\text{D},k} \ge 0\text{, }\forall k, \\ \label{MSRQ4}
&\;\;\;\;\;\;\sum\limits_{k \in \mathcal{K}} {\mathchar'26\mkern-10mu\lambda _\theta }\left( {{\bm{Q}}_{\text{D},k}{{\bm{T}}_l}} \right)  R_{\text{D},k} \le C_{\text{D},l}, \forall l, \forall k,\\ \label{MSRQ5}
&\;\;\;\;\;\;\left( {{2^{R_{{\text{D},\text{min}},k}}} - 1} \right){{\tilde{\gamma} }}_{\text{D},k}-{{\bm{h}}^{\rm H}_{\text{D},k}}{\bm{Q}}_{\text{D},k}{\bm{h}}_{\text{D},k}  \le 0{\rm{ }}, \forall k, \\ \label{MSRQ6}
&\;\;\;\;\;\;\left( {{2^{R_{{\text{U},\text{min}},j}}} - 1} \right){{\tilde{\gamma}_{\text{U},j}}}-{{P_{\text{U},j}{{\left| {{{ {{\bm{u}}^{\rm H}_{\text{U},j,z}} }}{\bm{h}}_{\text{U},j,z}} \right|}^2}}} \le 0{\rm{ }}, \forall j  \\ \label{MSRQ3}
& \;\;\;\;\;\;\text{and}\quad \eqref{MSR2}, \text{ }\eqref{MSR6}, \nonumber
\end{align}
\end{subequations}
where
\begin{subequations}\label{MSRQA}
\begin{align}
{\mathchar'26\mkern-10mu\lambda _\theta }&\left( {{\bm{Q}}_{\text{D},k}{{\bm{T}}_l}} \right)=1- {e^{ - \theta {\rm{Tr}}\left( {{\bm{Q}}_{\text{D},k}{{\bm{T}}_l}} \right)}},\\
\tilde {{\gamma }}_{\text{D},k}= &{\sum\limits_{{k \in \mathcal{K}}, k' \ne k} {{{{\bm{h}}^{\rm H}_{\text{D},k}}}{\bm{Q}}_{\text{D},k'}{\bm{h}}_{\text{D},k}}  + \sum\limits_{j \in \mathcal{J}} {P_{\text{U},j}{{\left| {{{h}}_{\text{IUI},{j,k}}} \right|}^2} + } \sigma _{\text{D},k}^2}, \\
{{\tilde{\gamma}_{\text{U},j}}}=&\sum\limits_{{j \in \mathcal{J}}, j' \ne j} {P_{\text{U},j'}{{\left| {{{ {{\bm{u}}^{\rm H}_{\text{U},j,z}} }}{\bm{h}}_{\text{U},j',z}} \right|}^2}}  + \sigma _{{\rm{U}},z}^2{\left\| {{{\bm{u}}_{{\rm{U}},j,z}}} \right\|^2} \\
 &   +{{\left\| {{{\bm{u}}_{{\rm{U}},j,z}}} \right\|}^2} \sum\limits_{i \in \mathcal{K}} {\sum\limits_{l \in \mathcal{L}} {{\rm{Tr}}\left({{\bm{Q}}_{{\rm{D}},k}}{{\bm{T}}_l}\right)} \delta _{{\rm{IRI}},l,z}^2}.  \nonumber
\end{align}
\end{subequations}


It is obvious that constraints \eqref{MSR2}, \eqref{MSR6}, \eqref{MSRQ1}, \eqref{MSRQ2}, \eqref{MSRQ5}, and \eqref{MSRQ6} are all of convex form due to the linear approximation. We only need to deal with constraint \eqref{MSRQ4} and the objective function.

Now, we consider converting the objective function to a convex form. Since the rate function can be expressed as a difference of two concave functions, i.e.,
\begin{equation}\label{MSRQfhR}
\begin{aligned}
{\sum\limits_{k \in \mathcal{K}} {\tilde {R}_{\text{D},k}} + \sum\limits_{j \in \mathcal{J}}{\tilde {R}_{\text{U},j}} } =f\left( {\bm{Q},\bm{P}} \right)-h\left( {\bm{Q},\bm{P}} \right),
\end{aligned}
\end{equation}
where
\begin{subequations}\label{MSRQf}
\begin{align}
f\left( {\bm{Q},\bm{P}} \right){\rm{ = }}&\sum\limits_{k \in \mathcal{K}} {{\rm{log_2}}\Big( { {{{ {{\bm{h}}^{\rm H}_{\text{D},k}} }}\bm{Q}_{\text{D},k}{{ {{\bm{h}}_{\text{D},k}} }}}} }
+ \tilde {{\gamma }}_{\text{D},k}  \Big) \nonumber\\
&{\rm{  + }}\sum\limits_{j \in \mathcal{J}} {{\rm{log_2}}\Big( { {P_{\text{U},j}{{\left| {{{ {{\bm{u}}^{\rm H}_{\text{U},j,z}} }}{\bm{h}}_{\text{U},j,z}} \right|}^2}}}}
 + \tilde{\gamma}_{\text{U},j}   \Big)\\
h\left( {\bm{Q},\bm{P}} \right){\rm{ = }}&\sum\limits_{k \in \mathcal{K}} {{\rm{log_2}}{\tilde{\gamma }}_{\text{D},k}}{\rm{ + }}\sum\limits_{j \in \mathcal{J}} {{\rm{log_2}}{{\tilde{\gamma}_{\text{U},j}}}},
\end{align}
\end{subequations}
we rewrite \eqref{MSRQ} as
\begin{equation}\label{MSRQfh}
\begin{aligned}
{\mathop {\max }\limits_{{\bm{w}}_{\text{D},k},{P}_{\text{U},j}} }\rm{}&f\left( {\bm{Q},\bm{P}} \right)-h\left( {\bm{Q},\bm{P}} \right)\\
{\rm s.t.}\quad &\eqref{MSRQ1}, \eqref{MSRQ2}, \eqref{MSR2}, \eqref{MSR6}, \eqref{MSRQ4}, \eqref{MSRQ5}, \eqref{MSRQ6}.
\end{aligned}
\end{equation}
It can be observed that \eqref{MSRQfhR} is a standard difference between two convex functions \cite{kha2012fast} program, and
the main work left for us now is determining how to convert concave function $h\left( {\bm{Q},\bm{P}} \right){\rm{  }}$ to make the objective function convex.
Inspired by \cite{kha2012fast}, we apply a first-order approximation of concave function $h\left( {\bm{Q},\bm{P}} \right)$.

Using ${\nabla _a}{\rm{ln}}\left( {{{b + c}}a} \right) = c{\left( {b + ca} \right)^{ - 1}}$,
we have the  inequality: ${\rm{log_2}}\left( {{{b + c}}a} \right) \le {\rm{log_2}}\left( {{{b + c}}{a_0}} \right) + c\frac{1}{{\ln 2}}{\left( {b + {{c}}{a_0}} \right)^{ - 1}}\left( {a - {a_0}} \right)$,
for $a \ge 0$. Then, the approximation is given in \eqref{MSRQh} at the top of the next page.
\begin{figure*}
\begin{equation}\label{MSRQh}
\begin{aligned}
 {h^{\left( n \right)}}\left( {{\bm{Q}},{\bm{P}}} \right) =& h\left( {{{\bm{Q}}^{\left( n \right)}},{{\bm{P}}^{\left( n \right)}}} \right) + \frac{\varphi _{\rm{D}}^{\left( n \right)}}{{\ln 2}}\sum\limits_{k \in \mathcal{K}} {\sum\limits_{j \in \mathcal{J}} {\left( {{P_{{\rm{U}},j}} - P_{{\rm{U}},j}^{\left( n \right)}} \right){{\left| {{{h}}_{\text{IUI},{j,k}}} \right|}^2}} } + \frac{{\phi _{\rm{U}}^{\left( n \right)}}}{{\ln 2}}\sum\limits_{j \in \mathcal{J}} {\left\| {{{\bm{u}}_{{\rm{U}},j,z}}} \right\|^2}{\sum\limits_{k \in \mathcal{K}}  } \sum\limits_{l \in \mathcal{L}} {{\rm{Tr}}\left({\bm{Q}_{{\rm{D},}k}} - {\bm{Q}}_{{\rm{D,}}k}^{\left( n \right)}\right){{\bm{T}}_l}} \delta _{{\rm{IRI}},l,z}^2\\
 &+ \frac{\varphi _{\rm{D}}^{\left( n \right)}}{{\ln 2}}\sum\limits_{k \in \mathcal{K}} {\sum\limits_{{k' \in \mathcal{K}}, k' \ne k} {{\bm{h}}_{{\rm{D}},k}^{\rm H}\left( {{{\bm{Q}}_{{\rm{D}},k'}} - {\bm{Q}}_{{\rm{D}},k'}^{\left( n \right)}} \right){{\bm{h}}_{{\rm{D}},k}}} }  + \frac{{\phi _{\rm{U}}^{\left( n \right)}} }{{\ln 2}}\sum\limits_{j \in \mathcal{J}} \sum\limits_{{j' \in \mathcal{J}}, j' \ne j} {\left( {{P_{{\rm{U}},j'}} - P_{{\rm{U}},j'}^{\left( n \right)}} \right){{\left| {{\bm{u}}_{{\rm{U}},j,z}^{\rm H}{{\bm{h}}_{{\rm{U}},j',z}}} \right|}^2}} \\
\end{aligned}
\end{equation}
\hrule
\end{figure*}
where $ {\varphi^{\left( {n} \right)}_\text{D}}$ and ${ \phi^{\left( {n} \right)} _\text{U}}$ are defined as
\begin{subequations}\label{MSRQvarphi}
\begin{align}
{\varphi^{\left( {n} \right)}_\text{D}} =&{\Big( {\sum\limits_{{k' \in \mathcal{K}}, k' \ne k} {{{ {{\bm{h}}_{\text{D},k}^{\rm H}}}}{{ {\bm{Q}^{\left( {n} \right)}_{\text{D},{k'}}} }}{{ {{\bm{h}}_{\text{D},k}} }}}  + \sum\limits_{j \in \mathcal{J}} {{{ {P^{\left( {n} \right)}_{\text{U},j}} }}{{\left| {{{h}}_{\text{IUI},{j,k}}} \right|}^2} + } \sigma _{\text{D},k}^2} \Big)^{-1}}\\
\phi _\text{U}^{\left( n \right)} =&{\Big(\sum\limits_{{j \in \mathcal{J}}, j' \ne j} {P_{{\rm{U}},j'}^{\left( n \right)}{{\left| {{\bm{u}}_{{\rm{U}},j,z}^{\rm H}{{\bm{h}}_{{\rm{U}},j',z}}} \right|}^2}}  + \sigma _{{\rm{U}},z}^2{\left\| {{{\bm{u}}_{{\rm{U}},j,z}}} \right\|^2}} \nonumber\\
 &+ \sum\limits_{k' \in \mathcal{K}} {\sum\limits_{l \in \mathcal{L}} {{\rm{Tr}}({\bm{Q}}_{{\rm{D}},k'}^{\left( n \right)}{{\bm{T}}_l})} \delta _{{\rm{IRI}},l,z}^2{{\left\| {{{\bm{u}}_{{\rm{U}},j,z}}} \right\|}^2}} \Big)^{ - 1}.
\end{align}
\end{subequations}

Note that, here, $\left({{\bm{Q}^{\left( {n} \right)}},{\bm{P}^{\left( {n} \right)}}}\right)$ denotes the value of $\left( {{\bm{Q}},{\bm{P}}}\right)$ at iteration $n$. Replacing $ h\left( {\bm{Q},\bm{P}} \right){\rm{  }}$ by its affine majorization in the neighbourhood of $\left({{\bm{Q}^{\left( {n} \right)}},{\bm{P}^{\left( {n} \right)}}}\right)$, we finally increase the objective in the next iteration. 
Subsequently, we deal with the downlink backhaul constraint \eqref{MSRQ4}. By introducing an auxiliary variable $\rho_{\text{D},k} \ge 0$, we can obtain
\begin{subequations}\label{BackhaulBCDapproximate}
\begin{align}
\sum\limits_{k \in \mathcal{K}} {\rho _{{\rm{D}},k}}\left( {1 - {e^{ - \theta {\rm{Tr}}\left( {{{\bm{Q}}_{{\rm{D}},k}}{{\bm{T}}_l}} \right)}}} \right) \le {C_{{\rm{D}},l}},   \forall l, \label{BackhaulBCDapproximate1}\\
R_{\text{D},k} \le \rho_{\text{D},k}, \forall k. \label{BackhaulBCDapproximate2}
\end{align}
\end{subequations}

As a result, $\rho_{\text{D},k}$ can be interpreted as the upper bound rate
constraint for downlink user $k$. Here, we can see that both \eqref{BackhaulBCDapproximate1} and \eqref{BackhaulBCDapproximate2} are of non-convex form, and then, we will approximate the two inequations one by one. First, we approximate inequation \eqref{BackhaulBCDapproximate1} as
\begin{subequations}\label{BackhaulBCDapproximatefurther}
\begin{align}
{\rho _{{\rm{D}},k}}\left( {1 - {e^{ - \theta {\rm{Tr}}\left( {{{\bm{Q}}_{{\rm{D}},k}}{{\bm{T}}_l}} \right)}}} \right) &\le {{\hat \rho }_{{\rm{D}},l,k}}, \label{BackhaulBCDapproximatefurther1}\\
\sum\limits_{k \in \mathcal{K}} {{{\hat \rho }_{\text{D},l,k}}}  &\le {C_{{\rm{D}},l}},  \label{BackhaulBCDapproximatefurther2}
\end{align}
\end{subequations}
where ${{{\hat \rho }_{{\rm{D}},l,k}}}$ is an auxiliary variable. By taking the natural logarithm of the left-hand-side (LHS) and RHS of inequality constraint \eqref{BackhaulBCDapproximatefurther1}, we rewrite \eqref{BackhaulBCDapproximatefurther1} as
\begin{equation}\label{BackhaulBCDLog}
\begin{aligned}
\log \left( {{\rho _{{\rm{D}},k}}} \right) + \log \left( {1 - {e^{ - \theta {\rm{Tr}}\left( {{{\bm{Q}}_{{\rm{D}},k}}{{\bm{T}}_l}} \right)}}} \right) \le \log \left( {{{\hat \rho }_{{\rm{D}},l,k}}} \right).
\end{aligned}
\end{equation}

It can be shown that $\log \left( {{{\hat \rho }_{{\rm{D}},l,k}}}\right)$  is a concave function over ${{\hat \rho }_{{\rm{D}},l,k}}$'s. However, the LHS of constraint \eqref{BackhaulBCDLog} is still
non-convex. Since $\log \left( {{\rho _{{\rm{D}},k}}} \right)$  and $\log \left( {1 - {e^{ - \theta {\rm{Tr}}\left( {{\bm{Q}_{{\rm{D}},k}}{{\bm{T}}_l}} \right)}}} \right)$ are concave functions over
${\rho _{{\rm{D}},k}}$ and ${\rm{Tr}}\left( {{{\bm{Q}}_{{{\rm{D}}},k}}{{\bm{T}}_l}} \right) $, respectively, their first-order approximations serve as their upper bounds.  Specifically, given any ${\rho ^{\left( n \right)}_{{\rm{D}},k}}$ and ${\rm{Tr}}\big( {{{\bm{Q}}^{\left( n \right)}_{{\rm{D}},k}}{{\bm{T}}_l}} \big) $, the first-order approximations of $\log \left( {{\rho _{{\rm{D}},k}}} \right)$  and $\log \left( {1 - {e^{ - \theta {\rm{Tr}}\left( {{{\bm{Q}}_{{\rm{D}},k}}{{\bm{T}}_l}} \right)}}} \right)$  can be expressed as
\begin{subequations}\label{BackhaulBCDLogapproximate}
\begin{align}
\log \left( {1 - {e^{ - \theta {\rm{Tr}}\left( {{{\bm{Q}}_{{\rm{D}},k}}{{\bm{T}}_l}} \right)}}} \right) \le&\Psi \left( {\bm{Q}}_{{\rm{D}},k}^{\left( n \right)} \right),\label{BackhaulBCDLogapproximate1}\\
\log \left( {{\rho _{{\rm{D}},k}}} \right) \le \frac{{{\rho _{{\rm{D}},k}} - \rho _{{\rm{D}},k}^{\left( n \right)}}}{{\rho _{{\rm{D}},k}^{\left( n \right)}}} &+ \log \left( {\rho _{{\rm{D}},k}^{\left( n \right)}} \right),&\label{BackhaulBCDLogapproximate2}
\end{align}
\end{subequations}
where
\begin{equation}\label{BackhaulBCDWhere}
\begin{aligned}
&\Psi \left( {\bm{Q}}_{{\rm{D}},k}^{\left( n \right)} \right)=\log \left( {1 - {e^{ - \theta {\rm{Tr}}\left( {{\bm{Q}}_{{\rm{D}},k}^{\left( n \right)}{{\bm{T}}_l}} \right)}}} \right)\\
&+\frac{{\theta {e^{ - \theta {\rm{Tr}}\left( {{\bm{Q}}_{{\rm{D}},k}^{\left( n \right)}{{\bm{T}}_l}} \right)}}{\rm{Tr}}\left( {{{\bm{Q}}_{{\rm{D}},k}}{{\bm{T}}_l}} \right)}}{{1 - {e^{ - \theta {\rm{Tr}}\left( {{\bm{Q}}_{{\rm{D}},k}^{\left( n \right)}{{\bm{T}}_l}} \right)}}}} - \frac{{\theta {e^{ - \theta {\rm{Tr}}\left( {{\bm{Q}}_{{\rm{D}},k}^{\left( n \right)}{{\bm{T}}_l}} \right)}}{\rm{Tr}}\left( {{\bm{Q}}_{{\rm{D}},k}^{\left( n \right)}{{\bm{T}}_l}} \right)}}{{1 - {e^{ - \theta {\rm{Tr}}\left( {{\bm{Q}}_{{\rm{D}},k}^{\left( n \right)}{{\bm{T}}_l}} \right)}}}} \nonumber
\end{aligned}
\end{equation}
and the equalities hold if and only if ${\bm{Q}}_{{\rm{D}},k}^{\left( n \right)}={\bm{Q}}_{{\rm{D}},k}$ and ${{\rho _{{\rm{D}},k}}} ={\rho _{{\rm{D}},k}^{\left( n \right)}}$, respectively. With \eqref{BackhaulBCDLogapproximate1} and \eqref{BackhaulBCDLogapproximate2}, the non-convex constraint \eqref{BackhaulBCDLog} can be approximated by the following convex constraint:
\begin{equation}\label{BackhaulBCDLogapproximateoverall}
\begin{aligned}
\frac{{{\rho _{{\rm{D}},k}} - \rho _{{\rm{D}},k}^{\left( n \right)}}}{{\rho _{{\rm{D}},k}^{\left( n \right)}}} &+ \log \left( {\rho _{{\rm{D}},k}^{\left( n \right)}} \right) + 
+\Psi \left( {\bm{Q}}_{{\rm{D}},k}^{\left( n \right)} \right) \le \log \left( {{{\hat \rho }_{{\rm{D}},l,k}}} \right).
\end{aligned}
\end{equation}

For constraint \eqref{BackhaulBCDapproximate2}, $R_{\text{D},k}$ is non-convex. Then, by introducing a set of auxiliary variables ${\mu _{{\rm{D}},k}} \ge 0$, we can approximate \eqref{BackhaulBCDapproximate2} as
\begin{subequations}\label{BackhaulBCDRate}
\begin{align}
\tilde {{\gamma }}_{\text{D},k} &\ge \frac{1}{{{\mu _{{\rm{D}},k}}}},\label{BackhaulBCDRate1}\\
{\mu _{{\rm{D}},k}}{\bm{h}}_{{\rm{D}},k}^{\rm H}{{\bm{Q}}_{{\rm{D}},k}}{{\bm{h}}_{{\rm{D}},k}} &\le {2^{{\rho _{{\rm{D}},k}}}} - 1. \label{BackhaulBCDRate2}
\end{align}
\end{subequations}

Note that $t\big({\mu _{{\rm{D}},k}}, {\bm{h}}_{{\rm{D}},k}^{\rm H}{{\bm{Q}}_{{\rm{D}},k}}{{\bm{h}}_{{\rm{D}},k}}\big)={\mu _{{\rm{D}},k}}{\bm{h}}_{{\rm{D}},k}^{\rm H}{{\bm{Q}}_{{\rm{D}},k}}{{\bm{h}}_{{\rm{D}},k}}$
is neither a convex nor a concave function of $\mu _{{\rm{D}},k}$ and ${\bm{h}}_{{\rm{D}},k}^{\rm H}{{\bm{Q}}_{{\rm{D}},k}}{{\bm{h}}_{{\rm{D}},k}}$. Fortunately, inspired by \cite{tran2012fast,beck2010sequential}, we recall the following inequality:
\begin{equation}\label{DRelaxRT}
\begin{aligned}
&t\big({\mu _{{\rm{D}},k}}, {\bm{h}}_{{\rm{D}},k}^{\rm H}{{\bm{Q}}_{{\rm{D}},k}}{{\bm{h}}_{{\rm{D}},k}}\big) \le T\big({\mu _{{\rm{D}},k}}, {\bm{h}}_{{\rm{D}},k}^{\rm H}{{\bm{Q}}_{{\rm{D}},k}}{{\bm{h}}_{{\rm{D}},k}}, \phi^{\left( n \right)} _{\text{D},k} \big) \\
&\;\;\;\;\;\;\;\;\;\;\;\;\;\;\;\;\;\;\;\;\;\;\;\;\;=\frac{1}{{2{ {\phi^{\left( n \right)} _{\text{D},k}} }}}{ {\mu^2 _{\text{D},k}} } + \frac{{{ {\phi^{\left( n \right)} _{\text{D},k}} }}}{2}{
{\left({\bm{h}}_{{\rm{D}},k}^{\rm H}{{\bm{Q}}_{{\rm{D}},k}}{{\bm{h}}_{{\rm{D}},k}}\right)^2 }},
\end{aligned}
\end{equation}
which holds for  ${ {\phi^{\left( n \right)} _{\text{D},k}} } > 0$. Then, we can use $T\big({\mu _{{\rm{D}},k}}, {\bm{h}}_{{\rm{D}},k}^{\rm H}{{\bm{Q}}_{{\rm{D}},k}}{{\bm{h}}_{{\rm{D}},k}}, \phi^{\left( n \right)} _{\text{D},k} \big)$ as an approximation of $t\big({\mu _{{\rm{D}},k}}, {\bm{h}}_{{\rm{D}},k}^{\rm H}{{\bm{Q}}_{{\rm{D}},k}}{{\bm{h}}_{{\rm{D}},k}}\big)$.

Furthermore, the convergence of the proposed algorithm can be proven by the following properties:
\begin{equation}\label{DRelaxRNT}
\begin{aligned}
t\big({\mu _{{\rm{D}},k}}, {\bm{h}}_{{\rm{D}},k}^{\rm H}{{\bm{Q}}_{{\rm{D}},k}}{{\bm{h}}_{{\rm{D}},k}}\big) = T\big({\mu _{{\rm{D}},k}}, {\bm{h}}_{{\rm{D}},k}^{\rm H}{{\bm{Q}}_{{\rm{D}},k}}{{\bm{h}}_{{\rm{D}},k}}, \phi^{\left( n \right)} _{\text{D},k} \big),  \nonumber\\
\nabla t\big({\mu _{{\rm{D}},k}}, {\bm{h}}_{{\rm{D}},k}^{\rm H}{{\bm{Q}}_{{\rm{D}},k}}{{\bm{h}}_{{\rm{D}},k}}\big) = \nabla T\big({\mu _{{\rm{D}},k}}, {\bm{h}}_{{\rm{D}},k}^{\rm H}{{\bm{Q}}_{{\rm{D}},k}}{{\bm{h}}_{{\rm{D}},k}}, \phi^{\left( n \right)} _{\text{D},k} \big),\nonumber
\end{aligned}
\end{equation}
when ${ {\phi^{\left( n \right)} _{\text{D},k}} }={\raise0.7ex\hbox{${\mu _{{\rm{D}},k}}$} \!\mathord{\left/  {\vphantom {{\mu _{{\rm{D}},k}} {\bm{h}}_{{\rm{D}},k}^{\rm H}{{\bm{Q}}_{{\rm{D}},k}}{{\bm{h}}_{{\rm{D}},k}}}}\right.\kern-\nulldelimiterspace} \!\lower0.7ex\hbox{${\bm{h}}_{{\rm{D}},k}^{\rm H}{{\bm{Q}}_{{\rm{D}},k}}{{\bm{h}}_{{\rm{D}},k}}$}}$.
Therefore, by approximating the LHS of \eqref{BackhaulBCDRate2} according to \eqref{DRelaxRT} and implementing a first-order Taylor expansion on the RHS of \eqref{BackhaulBCDRate2},
constraint \eqref{BackhaulBCDRate2} can be reformulated as
\begin{equation}\label{URelaxSCAR}
\begin{aligned}
\frac{1}{{2{ {\phi^{\left( n \right)} _{\text{D},k}} }}}{ {\mu^2 _{\text{D},k}} } + \frac{{{ {\phi^{\left( n \right)} _{\text{D},k}} }}}{2}&{
{\left({\bm{h}}_{{\rm{D}},k}^{\rm H}{{\bm{Q}}_{{\rm{D}},k}}{{\bm{h}}_{{\rm{D}},k}}\right)^2 }} \\
&\le {2^{\tilde \rho _{{\rm{D}},k}^{\left( n \right)}}}\left( {{{\tilde \rho }_{{\rm{D}},k}} - \tilde \rho _{{\rm{D}},k}^{\left( n \right)} + 1} \right) - 1.
\end{aligned}
\end{equation}

Finally, at iteration $n+1$,  the optimization problem can be summarized as
\begin{equation}\label{MSRQfhn}
\begin{aligned}
\mathop {\max }\limits_{\left\{\mathcal{A} \right\}} & {{f}}\left( {{\bm{Q}},{\bm{P}}} \right){\rm{ - }}{{{h}}^{\left( {\rm{n}} \right)}}\left( {{\bm{Q}},{\bm{P}}} \right)\\
s.t.\quad&\eqref{MSRQ1},\text{ } \eqref{MSRQ2},\text{ }   \eqref{MSRQ5},\text{ } \eqref{MSRQ6},\text{ } \eqref{BackhaulBCDapproximatefurther2},\text{ }  \eqref{BackhaulBCDLogapproximateoverall}\\
& \eqref{MSR2},\text{ } \eqref{BackhaulBCDRate1},\text{ } \eqref{URelaxSCAR}.
\end{aligned}
\end{equation}
where ${\mathcal{A}} =\left\{ {\bm{Q}}_{{\rm{D}},k}, {P_{{\rm{U}},j}}, {\rho_{{\rm{D}},k}}, {\mu _{{\rm{D}},k}}, {{\hat \rho}_{\text{D},l,k}} \right\}$.

When $\left\{ {P_{\text{U},j}, {\bm{w}}_{\text{D},k}} \right\}$ are fixed,
we can obtain the MMSE receiver for uplink user $j$:
\begin{equation}\label{MMSE}
\begin{aligned}
{\bm{u}}_{\text{U},j,z} = {{\bm{\Sigma }}^{-1}_{\text{U},j,z}{{\bm{h}}_{{\rm{U}},j,z}}\sqrt {{P_{{\rm{U}},j}}} },
\end{aligned}
\end{equation}
where
\begin{equation}\label{MSEE}
\begin{aligned}
{{\bm{\Sigma }}_{{\rm{U}},j,z}} =& \sum\limits_{{j' \in \mathcal{J}}} {{P_{{\rm{U}},j'}}} {{\bm{h}}_{{\rm{U}},j',z}}{\bm{h}}_{{\rm{U}},j',z}^{\rm H} \\
&+ \sum\limits_{{k' \in \mathcal{K}}} {\sum\limits_{l \in \mathcal{L}} {{\rm{Tr}}({{\bm{Q}}_{{\rm{D}},k'}}{{\bm{T}}_l})} } \delta _{{\rm{IRI}},l,z}^2{{\bm{I}}_M} + \sigma_{{\rm{U}},z}^2{{\bm{I}}_M}. \nonumber
\end{aligned}
\end{equation}


Overall, the detailed algorithm to solve problem \eqref{MSRAS1} with the iterative SDR-BCD-based algorithm is summarized  in Algorithm \ref{alg1}.
\begin{algorithm}[htb]
\setcounter{algorithm}{0}
\caption{Solution to Problem \eqref{MSRAS1} with the Iterative SDR-BCD-based Algorithm.}
\label{alg1}
\begin{algorithmic}[1] 
\REQUIRE ~~\\ {initialization: $\left\{\mathcal{A}, \bm{u}_{\text{U},j,z} \right\}^{\left( {0} \right)}$.\\
 $\text{Set } {n} {\rm{ }} \to {\rm{ 0}} $.} 
\REPEAT
\STATE Solve \eqref{MSRQfhn} with fixed receiver ${ {{\bm{u}}^{\dag}_{\text{U},j,z}} }\to  \bm{u}^{(n)}_{\text{U},j,z} $, and denote the optimal solutions as  $\left( {{{\bm{Q}}^\dag },{{\bm{P}}^\dag },{{\bm{\mu}}^\dag },{{\bm{\rho}}^\dag },{{\bm{\hat \rho}}^\dag }} \right)$.
\STATE Update $ \left\{\mathcal{A}\text{, }{\phi _{\text{D},k}}
\right\}^{\left( n+1 \right)}
\to \left\{\mathcal{A} \text{, }{\rm{ }} \frac{{{\mu_{{\rm{D}},k}}}}{ {{{\bm h}_{{\rm{D}},k}^{\rm{H}}{{\bm Q}_{{\rm{D}},k}}{{\bm h}_{{\rm{D}},k}}} }}\right\}^{\dag}$.

\STATE Obtain ${ {{\bm{u}}^{\dag}_{\text{U},j,z}} }$ by \eqref{MMSE} with fixed ${ {{\bm{Q}}^{\left( {n+1 } \right)}_{\text{D},k}} }$ and ${ {{\bm{P}}^{\left( {n+1 } \right)}_{\text{U},j}} }$.
\STATE $\text{Set } {n} {\rm{ }} \to {{n+1}} $.
\UNTIL Convergence.
\RETURN The optimal solutions $ \left\{ \bm{u}^\dag_{\text{U},j,z}, \bm{Q}^\dag_{\text{D},k},   P^\dag_{\text{U},j}\right\}$
\end{algorithmic}
\end{algorithm}


\subsection{Stage \uppercase\expandafter{\romannumeral2}: Solution to Problem \eqref{MSRAS2} with the Iterative SDR-BCD-based Algorithm}
Given the user association and MMSE receiver $ \bm{u}^{\left( {n} \right)}_{\text{U},j} $ in problem \eqref{MSRAS2},
we can see that constraints \eqref{MSR1}, \text{ }\eqref{MSR2}, \text{ }\eqref{MSR4} and  \text{ }\eqref{MSR5} can be transformed to convex expressions according to the SDR approach in Stage \uppercase\expandafter{\romannumeral1},
i.e., \eqref{MSRQ1}, \eqref{MSRQ2},  \eqref{MSRQ5} and \eqref{MSRQ6}.

Then, under fixed ${\bm{u}}_{\text{U},j,z}$, problem \eqref{MSRAS2} can be rewritten as
\begin{equation}\label{MSRQfhnSx}
\begin{aligned}
&\mathop {\max }\limits_{\left\{ {\mathcal{A}_0} \right\} }f\left(  {{\bm{Q}},{\bm{P}}} \right)-h^{\left( {n} \right)}\left(  {{\bm{Q}},{\bm{P}}} \right)\\
&s.t.\;\;{{\rm{Tr}}\left( {{\bm{Q}}_{\text{D},k}{{\bm{T}}_l}} \right)}=0, \quad \forall \psi \left( {{{\bm{\bar Q}}_{\text{D},k}{{\bm{T}}_l}}} \right)=0, \forall l, \forall k,\\
&\;\;\;\;\;\;\;\eqref{MSR2},\text{ } \eqref{MSRQ1},\text{ } \eqref{MSRQ2},\text{ }  \eqref{MSRQ4},\text{ } \eqref{MSRQ5},\text{ } \eqref{MSRQ6},
\end{aligned}
\end{equation}
where ${\mathcal{A}_0} =\left\{ {\bm{Q}}_{{\rm{D}},k}, {P_{{\rm{U}},j}}, {\rho_{{\rm{D}},k}}, {\mu _{{\rm{D}},k}} \right\}$.

With user-RAU association, non-convex constraint \eqref{MSRQ4} can be reformulated as
\begin{equation}\label{MSRQfhnS}
\begin{aligned}
&\sum\limits_{k \in \mathcal{K}_{\text{D},{l,k}}}^{} {\rho_{\text{D},k}} \le C_{\text{D},l},\\
\end{aligned}
\end{equation}
where $\mathcal{K}_{\text{D},{l}}$ represents the set of the user-RAU associations with RAU $l$ and $\rho_{\text{D},k}$ is formulated in \eqref{BackhaulBCDapproximate2}.
Then, similar to Algorithm \ref{alg1}, problem \eqref{MSRAS2} can be  approximated by the convex problem as
\begin{equation}\label{MSRQfhnSn}
\begin{aligned}
&\mathop {\max }\limits_{\left\{ {\mathcal{A}_0} \right\}}f\left(  {{\bm{Q}},{\bm{P}}} \right)-h^{\left( {n} \right)}\left(  {{\bm{Q}},{\bm{P}}} \right)\\
&s.t.\;\;{{\rm{Tr}}\left( {{\bm{Q}}_{\text{D},k}{{\bm{T}}_l}} \right)}=0, \quad \forall \psi \left( {{{\bm{\bar Q}}_{\text{D},k}{{\bm{T}}_l}}} \right)=0, \forall l, \forall k,\\
&\quad\;\;\eqref{MSRQ1},\text{ } \eqref{MSRQ2},\text{ } \eqref{MSRQ5},\text{ } \eqref{MSRQ6},\text{ } \eqref{MSRQfhnS},\text{ } \eqref{BackhaulBCDRate1}, \text{ }\eqref{MSR2},\text{ } \eqref{URelaxSCAR},
\end{aligned}
\end{equation}
where $\bm{\bar Q}_{\text{D},k} =  {{\bm{\bar w}}_{\text{D},k}} { {{\bm{\bar w}}^{\rm H}_{\text{D},k}} }$. 

The iterative SDR-BCD-based algorithm of problem \eqref{MSRAS2} is summarized in Algorithm \ref{alg2}.

The overall two-stage algorithm of problem \eqref{MSR} is summarized in Algorithm \ref{alg3}.
\begin{proposition}
\label{Thm_1}
The proposed iterative SDR-BCD-based algorithm is guaranteed to converge to a stationary point of problem \eqref{MSRQ}.
\begin{proof}
Please refer to Appendix A.
\end{proof}
\end{proposition}

\begin{algorithm}[htb]
\caption{Solution to Problem \eqref{MSRAS2} with the Iterative SDR-BCD-based Algorithm.}
\label{alg2}
\begin{algorithmic}[1] 
\REQUIRE ~~\\ {initialization: $\left\{\mathcal{A}_0, \bm{u}_{\text{U},j,z} \right\}^{\left( {0} \right)}$.\\
 $\text{Set } {n} {\rm{ }} \to {\rm{ 0}} $.} 
\REPEAT
\STATE Solve \eqref{MSRQfhnSn} with fixed receiver ${ {{\bm{u}}^{\dag}_{\text{U},j,z}} }\to  \bm{u}^{(n)}_{\text{U},j,z} $,  and denote the optimal solutions as  $\left( {{{\bm{Q}}^\dag },{{\bm{P}}^\dag },{{\bm{\mu}}^\dag },{{\bm{\rho}}^\dag }} \right)$.
\STATE Run 3, 4 and 5 of Algorithm \ref{alg1}.
\UNTIL Convergence.
\RETURN The optimal solutions $ \left\{\bm{u}^\dag_{\text{U},j,z},   \bm{Q}^\dag_{\text{D},k},   P^\dag_{\text{U},j} \right\}$.
\end{algorithmic}
\end{algorithm}

\begin{algorithm}[htb]
\caption{ Overall Iterative SDR-BCD-based Algorithm.}
\label{alg3}
\begin{algorithmic}[1] 
\STATE Run Algorithm \ref{alg1} to obtain the DU-RAU association.
\STATE Run Algorithm \ref{alg2} to obtain the sparse beamforming and power control strategy.
\end{algorithmic}
\end{algorithm}


Problem \eqref{MSRQfhn} is a linear program (LP) that  has $2K{M^2}{L^2}+J$
real variables and $4J+4K+2L+KL$ linear constraints. The problem can be solved efficiently by interior point method that will take $O(\sqrt{2K{M^2}{L^2}+J}\text{log}(1/\epsilon))$  iterations, with each iteration requiring at most $O((2K{M^2}{L^2}+J)^3+(2K{M^2}{L^2}+J)(4J+4K+2L+KL))$ \cite{algorithms1997theory}, where $\epsilon$ is the precision requiring for solving the problem.

\section{The Proposed SPCA-based Algorithm}
In the preceding section, we needed to solve an SDP problem in each iteration. As is known, solving an SDP problem requires relatively high computational complexity.
In this section, we develop a low-complexity algorithm named the SPCA-based algorithm. Similar to Section \uppercase\expandafter{\romannumeral3}, we also formulate the SPCA-based  algorithm into two-stage optimization subproblems.

\subsection{Stage \uppercase\expandafter{\romannumeral1}: Solution to Problem \eqref{MSRAS1} with the Iterative SPCA-based Algorithm}
Motivated by \cite{tran2012fast}, we first reformulate problem \eqref{MSR} as
\begin{subequations}\label{ReForm}
\begin{align}
&{\mathop {\max }\limits_{{\bm{w}}_{\text{D},k},{\bm{u}}_{\text{U},j,z},{P}_{\text{U},j}} } \prod\limits_{k \in \mathcal{K}} {\left( {1 +
r_{\text{D},k}} \right)\prod\limits_{j \in \mathcal{J}} {\left( {1 +
r _{\text{U},j}} \right)} }  \label{ReForm1}\\
&s.t.\quad\left( {{2^{{R_{{\text{D},\text{min}},k}}}} - 1} \right) { \gamma _{\text{D},k}} \le {{{ {{\bm{w}}^{\rm H}_{\text{D},k}} }}{\bm{H}}_{\text{D},k}{\bm{w}}_{\text{D},k}}, \forall k,  \label{ReFor4}\\ \label{ReFor5}
&\quad\;\;\;\;\; \left( {{2^{{R_{{\text{U},\text{min}},j}}}} - 1} \right) { \gamma_{\text{U},j}} \le P_{\text{U},j}{\left| {{{ {{\bm{u}}^{\rm H}_{\text{U},j,z}} }}{\bm{h}}_{\text{U},j,z}} \right|^2} ,
\forall j\in \tilde{{\cal J}_{z}} \\ \label{ReFor6}
&\quad\;\;\;\;\; \eqref{MSR1},\text{ } \eqref{MSR2},\text{ } \eqref{MSR6},\text{ } \eqref{MSRAS11},
\end{align}
\end{subequations}
which can be further equivalently rewritten as
\begin{subequations}\label{Relax}
\begin{align}
{\mathop {\max }\limits_{{\bm{w}}_{\text{D},k},{\bm{u}}_{\text{U},j,z},{P}_{\text{U},j}} } & \prod\limits_{k \in \mathcal{K}} {\chi _{\text{D},k}\prod\limits_{j \in \mathcal{J}} {\chi _{\text{U},j}} } \\ \label{Relax1}
s.t.\quad&1 + { r} _{\text{D},k} \ge \chi _{\text{D},k}, \forall k,\\ \label{Relax2}
&1 + { r} _{\text{U},j} \ge \chi _{\text{U},j}, \forall j,  \\ \label{Relax3}
&\chi _{\text{U},j} \ge 1,\chi _{\text{D},k} \ge 1, \forall k, \forall j, \\ \label{Relax4}
&\eqref{MSR1},\text{ } \eqref{MSR2},\text{ } \eqref{MSR6},\text{ }  \eqref{MSRAS11},\text{ } \eqref{ReFor4}, \text{ }\eqref{ReFor5},
\end{align}
\end{subequations}
where ${{\bm{H}}_{\text{D},k} = {\bm{h}}_{\text{D},k}{{{\bm{h}}^{\rm H}_{\text{D},k}}}}$.

Note that the objective function in \eqref{Relax} admits an SOC representation  \cite{tran2012fast,lobo1998applications}, as shown at the top of the next page. In particular, we rewrite the objective function in \eqref{Soc} as
\begin{figure*}
\begin{subequations}\label{Soc}
\begin{align}
{\mathop {\max }\limits_{{\bm{w}}_{\text{D},k},{\bm{u}}_{\text{U},j,z},{P}_{\text{U},j}} }  &\text{ } {\rm{ }}{\vartheta ^{\left( 0 \right)}}\\
s.t.\quad&\text{ }{\rm{ }}{\left\| {\left[ {2\vartheta _k^{\left( {N - 1} \right)}{\rm{    }}\left( \chi_{x,2k - 1} - {\chi _{x,2k}} \right)} \right]} \right\|_2} \le \left( \chi _{x,2k - 1} + {\chi _{x,2k}} \right),{\rm{  }}  \text{ }k = 1, \cdots ,{2^{N - 1}}, \label{Soc1}\\
&\text{ }{\left\| {\left[ {2\vartheta _{\tilde k }^{\left( t \right)}{\rm{    }}\left( {\vartheta _{\tilde k  - 1}^{\left( {t + 1} \right)} - \vartheta _{\tilde k }
^{\left( {t + 1} \right)}} \right)} \right]} \right\|_2} \le \left( {\vartheta _{\tilde {k}  - 1}^{\left( {t + 1} \right)} + \vartheta _{\tilde k }^{\left( {t + 1} \right)}} \right),
\text{ }t = 0, \cdots ,\left( {N - 2} \right),\mathord{\tilde k}  = 1, \cdots ,{2^t}  \label{Soc2}
\end{align}
\end{subequations}
\hrule
\end{figure*}
where $N$ is some positive integer, and $N=\left\lceil {{\rm{log}}_2{{(K + J) }}} \right\rceil $, where
$\left\lceil {\bar x {\rm{ }}} \right\rceil $ is the smallest integer not less than $\bar x$.
$\chi_{x}$ is defined as
\begin{equation}\label{Other}
\begin{aligned}
{\chi _x} = \left\{ {\begin{array}{*{20}{c}}
{{\chi_\text{D}^{}}, x \in \mathcal{K}},\\
{{\chi_\text{U}^{}}, x \in \mathcal{J}}.
\end{array}} \right.
\end{aligned}
\end{equation}
It is obvious  that the order of $x$  will not affect the result when the following relevant expression is one-to-one matched. In practice, we first consider $ \chi _\text{D}$ and then $ \chi _\text{U}$. We note that the expression is satisfied if and only if ${{\rm{log}}_2{{(K + J) }}}=\left\lceil {{\rm{log}}_2{{(K + J) }}} \right\rceil $; when
${{\rm{log}}_2{{(K + J) }}} \ne \left\lceil {{\rm{log}}_2{{(K + J) }}} \right\rceil $, we define an additional $\chi_{x,x'}=1$ for $x'=K+J+1,\cdots, 2^N $.
Then, we consider the constraint of \eqref{Relax}. We can see that  \eqref{Relax3}, \eqref{MSR1} and \eqref{MSR2} are already convex forms. Thus, the next work of the paper is mainly to deal with the constraints in \eqref{Relax1}, \eqref{Relax2}, \eqref{MSRAS11},\text{ } \eqref{ReFor4} and \text{ }\eqref{ReFor5}.

To begin, we can rewrite constraint \eqref{Relax1}  as
\begin{equation}\label{DRelax}
\begin{aligned}
 \gamma _{\text{D},k} \le \frac{{{ {{\bm{w}}^{\rm H}_{\text{D},k}} }}{\bm{H}}_{\text{D},k}{\bm{w}}_{\text{D},k}}{{\chi _{\text{D},k} - 1}}.
\end{aligned}
\end{equation}
We observe that \eqref{DRelax} is non-convex. Since the RHS has the form of quadratic-over-linear,
it can be replaced by its first-order expansions \cite{boyd2004convex}. Thus, we can transform the problem into convex programming.
Specifically, we define the function: $g\left( {{\bm{w}},\chi,{\bm{A}} } \right)   = \frac{{{{\bm{w}}^{\rm H}}{\bm{A}} {\bm{w}} }}{{\chi  - a}}$,
where $\bm{A} \ge 0$ and $\chi  \ge a$. Then, we obtain the first-order Taylor expansion of $g\left( {{\bm{w}},\chi,{\bm{A}} } \right)$ about a certain point $\left( {{\bm{w}}^{\left( n \right)}}, {\chi ^{\left( n \right)}}  \right)$ as
\begin{equation}\label{FirstOrder}
\begin{aligned}
&G\left( {{\bm{w}},\chi ,{{\bm{w}}^{\left( n \right)}},{\chi ^{\left( n \right)}},{\bm{A}}},a \right)\\
 &= \frac{{2\Re \left\{ {{{\left( {{{\bm{w}}^{\left( n \right)}}} \right)}^{\rm H}}{\bm{Aw}}} \right\}}}{{{\chi ^{\left( n \right)}} - a}}
- \frac{{{{\left( {{{\bm{w}}^{\left( n \right)}}} \right)}^{\rm H}}{\bm{A}}{{\bm{w}}^{\left( n \right)}}}}{{{{\left( {{\chi ^{\left( n \right)}} - a} \right)}^2}}}\left( {\chi  - a} \right).
\end{aligned}
\end{equation}

By replacing the RHS of \eqref{DRelax} by \eqref{FirstOrder}, we can transform constraint \eqref{Relax1} into a convex form, which is
\begin{equation}\label{DRelaxR}
\begin{aligned}
 \gamma _{\text{D},k} \le G\left( {{\bm{w}}_{\text{D},k},\chi _{\text{D},k},{{ {{\bm{w}}^{\left( n \right)}_{\text{D},k}} }},{{ {\chi^{\left( n \right)} _{\text{D},k}} }},{\bm{H}}_{\text{D},k}},1 \right).
\end{aligned}
\end{equation}

Next, we note that the RHS of constraint  \eqref{ReFor4} has a similar expansion compared to \eqref{Relax1}. Therefore, similar to \eqref{FirstOrder}, we define the function: $s\left( {\bm{w}} \right) = {{\bm{w}}^{\rm H}}{\bm{Aw}}$, and obtain the following first-order Taylor expansion as
\begin{equation}\label{ratezUx}
\begin{aligned}
S\left( {{\bm{w}},{{\bm{w}}^{\left( n \right)}},{\bm{A}}} \right)=& {\Big( {{{\bm{w}}^{\left( n \right)}}} \Big)^{\rm H}}{\bm{A}}{{\bm{w}}^{\left( n \right)}} \\
& + 2\Re \left\{ {{{\Big( {{{\bm{w}}^{\left( n \right)}}} \Big)}^{\rm H}}{\bm{A}}} \Big( {{\bm{w}} - {{\bm{w}}^{\left( n \right)}}} \Big)\right\}.
\end{aligned}
\end{equation}

Thus, \eqref{ReFor4} can be approximated by
\begin{equation}\label{Qos}
\begin{aligned}
\left( {{2^{{R_{{\text{D},\text{min}},k}}}} - 1} \right)\gamma _{\text{D},k} \le S\left( {{{{{\bm{w}}_{\text{D},{k}}}},{{{\bm{w}}^{\left( n \right)}_{\text{D},{k}}}},{\bm{H}}_{\text{D},k}}} \right).
\end{aligned}
\end{equation}

Next, we deal with constraints \eqref{ReFor5} and \eqref{Relax2}. To facilitate the analysis, we first rewrite \eqref{Relax2} as
\begin{equation}\label{URelax}
\begin{aligned}
\gamma _{\text{U},j} \le \frac{{P_{\text{U},j}{{\left| {{{ {{\bm{u}}^{\rm H}_{\text{U},j,z}} }}{\bm{h}}_{\text{U},j,z}} \right|}^2}}}{{\chi _{\text{U},j} - 1}}.
\end{aligned}
\end{equation}
Since both constraints \eqref{ReFor5} and \eqref{URelax}  share the same  variables, i.e., $\gamma _{\text{U},j}$ and $P_{\text{U},j}{\left| {{{ {{\bm{u}}^{\rm H}_{\text{U},j,z}} }}{\bm{h}}_{\text{U},j,z}} \right|^2}$, we will deal with the two constraints in a similar way. To begin with, we first consider the LHS variable of both constraints, i.e., $\gamma _{\text{U},j}$.
It can be observed that $\gamma _{\text{U},j}$ involves a quartic term of the optimization variables, i.e.,
\begin{equation}\label{URelaxz}
\begin{aligned}
{\gamma_{{\rm{U}},{\rm{D}},j,z}} = \sum\limits_{k \in \mathcal{K}} { {\sum\limits_{l \in \mathcal{L}} {{{\left\| {{{\bm{w}}_{{\rm{D}},l,k}}} \right\|}^2}} \delta _{{\rm{IRI}},l,z}^2} } {\left\| {{{\bm{u}}_{{\rm{U}},j,z}}} \right\|^2},
\end{aligned}
\end{equation}
which is the most difficult part. By introducing a series of variables ${{\tilde P}_{{\rm{D}},l}}$ and ${\bar {\tilde P}_{{\rm{D}},l,j}}$,
we approximate $\gamma _{\text{U},\text{D},j,z}$ as
 \begin{subequations}\label{DWMMSEE}
\begin{align}
&\mathop \sum \limits_{k \in \mathcal{K}} {\left\| {{{\bm{w}}_{{\rm{D}},l,k}}} \right\|^2} \le {{\tilde P}_{{\rm{D}},l}}, \label{DWMMSEE1}\\
& {{{\tilde P}_{{\rm{D}},l}}\delta _{{\rm{IRI}},l,z}^2} {\left\| {{{\bm{u}}_{{\rm{U}},j,z}}} \right\|^2} \le \bar {\tilde P}_{{\rm{D}},l,j}^2, \label{DWMMSEE2}\\
&0 \le {{\tilde P}_{{\rm{D}},l}} \le {\rm{\bar P}}_{\text{D},l}.  \label{DWMMSEE4}
\end{align}
\end{subequations}
However, the inequality \eqref{DWMMSEE2} is still non-convex. To proceed, we introduce the following lemma:
\begin{lemma}
\label{Lem_0}
Let $a>0$ and $b \in {\mathbb {{R}}}$ be some arbitrary variables, $\bm{d} \in {\mathbb {{C}}}^{\bar N}$ be a vector, $c$ be a constant value.
Then, $ac{\left\| \bm{d} \right\|^2} \le {b^2}$ can be approximated by the following inequality:
\begin{equation}\label{DRelaxRNTX}
\begin{aligned}
{\left( {{a^{\left( n \right)}}} \right)^2}c{\left\| \bm{d} \right\|^2} - 2b{a^{\left( n \right)}}{b^{\left( n \right)}} + {\left( {{b^{\left( n \right)}}} \right)^2}a \le 0,
\end{aligned}
\end{equation}
where ${a^{\left( n \right)}}$ and ${b^{\left( n \right)}}$ are the corresponding feasible points of $a$ and $b$, respectively.
\begin{proof}
First, we rewrite the inequality as
\begin{equation}\label{DRelaxRNTXP}
\begin{aligned}
c{\left\| \bm{d} \right\|^2} - \frac{{{b^2}}}{a} \le 0.
\end{aligned}
\end{equation}
Then, by approximating $\frac{{{b^2}}}{a}$ with its first-order Taylor expansion, we can obtain the desired result.
\end{proof}
\end{lemma}

Further, according to Lemma \ref{Lem_0}, we can see that \eqref{DWMMSEE2}
can be approximated by the following convex constraint:
\begin{equation}\label{DWMMSEF}
\begin{aligned}
{\Big( {\tilde P_{{\rm{D}},l}^{\left( n \right)}} \Big)^2}&{\delta _{{\rm{IRI}},l,z}^2} {\left\| {{{\bm{u}}_{{\rm{U}},j,z}}} \right\|^2} \\
&- 2{{\bar {\tilde P}}_{{\rm{D}},l,j}}{\bar {\tilde P}}_{{\rm{D}},l,j}^{\left( n \right)}\tilde P_{{\rm{D}},l}^{\left( n \right)} + {\Big( {\bar {\tilde P}_{{\rm{D}},l,j}^{\left( n \right)}} \Big)^2}{{\tilde P}_{{\rm{D}},l}} \le 0,
\end{aligned}
\end{equation}

Although we convert the quartic term of $\gamma _{\text{U},\text{D},j,z}$ into convex form, there still exists a non-convex expression in $\gamma _{\text{U},j}$, that is, $\sum\limits_{{j \in \mathcal{J}},  j' \ne j} {P_{\text{U},j'}{{\left| {{{ {{\bm{u}}^{\rm H}_{\text{U},j,z}} }}{\bm{h}}_{\text{U},j',z}} \right|}^2}} $.
Motivated by \cite{zhou2018energy}, we obtain
\begin{equation}\label{Power}
\begin{aligned}
&{\left| {{{ {{\bm{u}}^{\rm H}_{\text{U},j,z}} }}{\bm{h}}_{\text{U},j',z}} \right|^2} \le \frac{{{{ {\beta^2 _{\text{U},{j,j'}}} }}}}{{P_{\text{U},j'}}},
\end{aligned}
\end{equation}
where $\beta _{\text{U},{j,j'}}$ is a newly introduced variable.

Then, by applying  Lemma \ref{Lem_0},  \eqref{Power} can be approximated as
\begin{small}
\begin{equation}\label{FirstOrderCL}
\begin{aligned}
{\Big( P^{\left( n \right)}_{\text{U},j'}\Big)^2}&{\left| {{{ {{\bm{u}}^{\rm H}_{\text{U},j,z}} }}{\bm{h}}_{\text{U},j',z}} \right|^2} \\
&- 2\beta _{\text{U},{j,j'}}{\beta ^{\left( n \right)}_{\text{U},{j,j'}}}P^{\left( n \right)}_{\text{U},j'} + {\Big( {\beta ^{\left( n \right)}_{\text{U},{j,j'}}} \Big)^2}P_{\text{U},j'} \le 0.
\end{aligned}
\end{equation}
\end{small}

Then, according to the above approximation, we can represent \eqref{ReFor5} as
\begin{equation}\label{FirstOrderCLResort}
\begin{aligned}
\left( {{2^{{R_{{\rm{U}},{\rm{min}},j}}}} - 1} \right)\bar{\gamma}_{\text{U},j} \le {\left| {{\bm{u}}_{{\rm{U}},j,z}^{\rm H}{{\bm{h}}_{{\rm{U}},j,z}}} \right|^2},
\end{aligned}
\end{equation}
where
\begin{equation}\label{FirstOrderCLWhere}
\begin{aligned}
\bar{\gamma}_{\text{U},j}= {\sum\limits_{{j \in \mathcal{J}}, j' \ne j} {\frac{{\beta _{{\rm{U}},j,j'}^2}}{{{P_{{\rm{U}},j}}}}}  + \frac{{\sigma _{U,z}^2{{\left\| {{{\bm{u}}_{{\rm{U}},j,z}}} \right\|}^2}}}{{{P_{{\rm{U}},j}}}} + \sum\limits_{l \in \mathcal{L}} {\frac{{\bar {\tilde P}_{{\rm{D}},l,j}^2}}{{{P_{{\rm{U}},j}}}}} }.
\end{aligned}
\end{equation}

It is obvious that the LHS of \eqref{FirstOrderCLResort} is converted to convex form, and similar to \eqref{ratezUx}, we approximate the RHS of \eqref{FirstOrderCLResort} and  rewrite \eqref{FirstOrderCLResort} as
\begin{equation}\label{FirstOrderCLResortrewrite}
\begin{aligned}
\left( {{2^{{R_{{\rm{U}},{\rm{min}},j}}}} - 1} \right)\bar{\gamma}_{\text{U},j}\le S\left( {{\bm{u}}_{{\rm{U}},j,z},{{\bm{u}}_{{\rm{U}},j,z}^{\left( n \right)}},{\bm{H}}_{{\rm{U}},j,z}} \right),
\end{aligned}
\end{equation}
where ${\bm{H}}_{{\rm{U}},j,z}={{\bm{h}}_{{\rm{U}},j,z}}{{\bm{h}}^{\rm H}_{{\rm{U}},j,z}}$.

Here, \eqref{URelax} is still non-convex, as there is a variable in the denominator on the RHS of the inequality.
Similar to \eqref{FirstOrderCLResort}, \eqref{URelax} can be equivalently represented as
\begin{equation}\label{URelaxUPbound}
\begin{aligned}
\bar{\gamma}_{\text{U},j} \le \frac{{{{\left| {{{ {{\bm{u}}^{\rm H}_{\text{U},j,z}} }}{\bm{h}}_{\text{U},j,z}} \right|}^2}}}{{\chi _{\text{U},j} - 1}}.
\end{aligned}
\end{equation}
Additionally, we further approximate \eqref{URelaxUPbound} as
\begin{equation}\label{URelaxUPboundapproximate}
\begin{aligned}
\bar{\gamma}_{\text{U},j} \le G\left( {{\bm{u}}_{{\rm{U}},j,z},\chi ,{{\bm{u}}_{{\rm{U}},j,z}^{\left( n \right)}},{\chi ^{\left( n \right)}},{\bm{H}}_{{\rm{U}},j,z}},1 \right).
\end{aligned}
\end{equation}

Finally, we focus on the backhaul constraint \eqref{MSRAS11}. Similar to \eqref{BackhaulBCDapproximate}, by introducing an auxiliary variable $\tilde \rho_{\text{D},k} \ge 0$,
constraint \eqref{MSRAS11} can be approximated as
 \begin{subequations}\label{RelaxSCAB}
\begin{align}
\sum\limits_{k \in \mathcal{K}} {{{\tilde \rho }_{{\rm{D}},k}}\left( {1 - {e^{ - \theta \left\| { {{\bm{w}}_{{\rm{D}},l,k}^{}} } \right\|_2^2}}} \right)}  &\le {C_{{\rm{D}},l}},\label{RelaxSCAB1}\\
R_{\text{D},k} &\le \tilde \rho_{\text{D},k}.\label{RelaxSCAB2}
\end{align}
\end{subequations}

It is obvious that both \eqref{RelaxSCAB1} and \eqref{RelaxSCAB2} are in non-convex form, and in the following, we will first handle \eqref{RelaxSCAB1} and then \eqref{RelaxSCAB2}. To begin with, similar to
\eqref{BackhaulBCDLog}, we approximate \eqref{RelaxSCAB1} as
\begin{subequations}\label{RelaxSCABapproximate}
\begin{align}
{{\tilde \rho }_{{\rm{D}},k}}\left( {1 - {e^{ - \theta \left\| { {{\bm{w}}_{{\rm{D}},l,k}^{}} } \right\|_2^2}}} \right) \le \mathord{\bar {\tilde \rho} } _{{\rm{D}},k,l}^2,\label{RelaxSCABapproximate1}\\
\sum\limits_{k \in \mathcal{K}}   \mathord{\bar {\tilde \rho} } _{{\rm{D}},k,l}^2\le {C_{{\rm{D}},l}},\label{RelaxSCABapproximate2}
\end{align}
\end{subequations}
where $\mathord{\bar {\tilde \rho} } _{{\rm{D}},k,l}$ is a newly introduced variable.
Here, \eqref{RelaxSCABapproximate2} is of convex form, and then, we represent \eqref{RelaxSCABapproximate1} as
\begin{equation}\label{RelaxSCABapproximateresort}
\begin{aligned}
1 - {e^{ - \theta \left\| { {{\bm{w}}_{{\rm{D}},l,k}} } \right\|_2^2}} \le {\rm{ }}\frac{{\bar{ \tilde \rho} _{{\rm{D}},k,l}^2}}{{{{\tilde \rho }_{{\rm{D}},k}}}}.
\end{aligned}
\end{equation}

Since $1 - {e^{ - \theta {\bm{w}}}}$ is convex function over ${\bm{w}}$, its first-order approximation serves as its upper bound. Specifically, given any ${\bm{w}}^{\left( n \right)}$, the first-order approximation of $1 - {e^{ - \theta {\bm{w}}}}$ can be expressed as
\begin{equation}\label{RelaxSCABCfirorder}
\begin{aligned}
1 - {e^{ - \theta {\bm{w}}}} &= V\left( {{\bm{w}},{{\bm{w}}^{\left( n \right)}}} \right)\\
& =1 - {e^{ - \theta {\bm{w}}^{\left( n \right)}}} + \theta {e^{ - \theta {\bm{w}}^{\left( n \right)}}}\left( {{\bm{w}} - {\bm{w}}^{\left( n \right)}} \right).
\end{aligned}
\end{equation}

Then, by employing Lemma \ref{Lem_0} and \eqref{RelaxSCABCfirorder}, \eqref{RelaxSCABapproximateresort} can be approximated as
\begin{equation}\label{RelaxSCABapproximateall}
\begin{aligned}
&V\left( {\left\| { {{\bm{w}}_{{\rm{D}},l,k}} } \right\|_2^2,\left\| { {{\bm{w}}_{{\rm{D}},l,k}^{\left( n \right)}} } \right\|_2^2} \right){\left( {\tilde \rho _{{\rm{D}},k}^{\left( n \right)}} \right)^2}\\
 &\;\;\;\;\;\;\;\;\;\;\;\;\;\;\;\;\;\;\le 2{{\bar {\tilde \rho }}_{{\rm{D}},k,l}}\bar {\tilde \rho} _{{\rm{D}},k,l}^{\left( n \right)}\tilde \rho _{{\rm{D}},k}^{\left( n \right)} - {\left( {\bar {\tilde \rho} _{{\rm{D}},k,l}^{\left( n \right)}} \right)^2}{{\tilde \rho }_{{\rm{D}},k}}.
\end{aligned}
\end{equation}

It can be seen that, by now, \eqref{RelaxSCAB1} has been approximated by a convex expression. Next, we will handle \eqref{RelaxSCAB2} by employing the newly introduced variable ${\tilde \mu } _{\text{D},k}$, and \eqref{RelaxSCAB2} can be approximated by the following expressions:
\begin{subequations}\label{RelaxSCABCnewvariable}
\begin{align}
\sum\limits_{{k' \in \mathcal{K}}, k' \ne k} {{\left| {{{ {{{\bm{h}}^{\rm H}_{\text{D},k}} }}}{\bm{w}}_{\text{D},k'}} \right|}^2}  &+ \sum\limits_{j \in \mathcal{J}} {p_{\text{U},j}{{\left| {{{h}}_{\text{IUI},{j,k}}} \right|}^2} + } \sigma _n^2 \ge \tilde \mu _{\text{D},k}, \label{RelaxSCABCnewvariable1}\\
 &\frac{{{{\left| {{\bm{h}}_{{\rm{D}},k}^{\rm H}{{\bm{w}}_{{\rm{D}},k}}} \right|}^2}}}{{{{\tilde \mu }_{{\rm{D}},k}}}} \le  {{2^{{{\tilde \rho }_{{\rm{D}},k}}}} - 1}. \label{RelaxSCABCnewvariable2}
\end{align}
\end{subequations}

Since ${{\left| {{{ {{{\bm{h}}^{\rm H}_{\text{D},k}} }}}{\bm{w}}_{\text{D},k'}} \right|}^2}$ and ${2^{{{\tilde \rho }_{{\rm{D}},k}}}}$  are concave functions over ${\bm{w}}_{\text{D},k'}$ and ${{\tilde \rho }_{{\rm{D}},k}}$, we can obtain the lower bound of them. Then, taking a further step, we have the following convex constraints to approximate
constraint \eqref{RelaxSCABCnewvariable1} and constraint \eqref{RelaxSCABCnewvariable2}:
\begin{subequations}\label{RelaxSCABCnewvariable1lowerbound}
\begin{align}
&\sum\limits_{{k' \in \mathcal{K}}, k' \ne k} S\left( {{{{{\bm{w}}_{\text{D},{k'}}}},{{{\bm{w}}^{\left( n \right)}_{\text{D},{k'}}}},{\bm{H}}_{\text{D},k}}} \right)  + \sum\limits_{j \in \mathcal{J}} {p_{\text{U},j}{{\left| {{{h}}_{\text{IUI},{j,k}}} \right|}^2} + } \sigma _n^2 \ge \tilde \mu _{\text{D},k}, \label{RelaxSCABCnewvariable1lowerbound1}\\
 &\frac{{{{\left| {{\bm{h}}_{{\rm{D}},k}^{\rm H}{{\bm{w}}_{{\rm{D}},k}}} \right|}^2}}}{{{{\tilde \mu }_{{\rm{D}},k}}}} \le  {2^{\tilde \rho _{{\rm{D}},k}^{\left( n \right)}}}\left( {{{\tilde \rho }_{{\rm{D}},k}} - \tilde \rho _{{\rm{D}},k}^{\left( n \right)} + 1} \right) - 1. \label{RelaxSCABCnewvariable1lowerbound2}
\end{align}
\end{subequations}

Then, the original problem \eqref{ReForm} can be reformulated as a series of the convex approximate problems discussed above. Iteration $n + 1$ of the proposed approach is
\begin{subequations}\label{RelaxA}
\begin{align}
\mathop {\max }\limits_{\left\{\mathcal{B} \right\}}
& \quad\quad{\vartheta ^{\left( 0 \right)}}\\ \label{RelaxA11}
s.t.\quad&\text{ }\eqref{MSR1},\text{ } \eqref{MSR2}, \text{ }\eqref{MSR6}, \text{ }  \eqref{Relax3}, \text{ }\eqref{Soc1},\text{ } \eqref{Soc2},\\
&\text{ }\eqref{DRelaxR}, \text{ }\eqref{Qos}, \text{ }\eqref{DWMMSEE1},  \text{ }\eqref{DWMMSEE4}, \text{ }\eqref{DWMMSEF}, \text{ }\eqref{FirstOrderCL},\\
&\text{ }\eqref{FirstOrderCLResortrewrite},  \text{ }\eqref{URelaxUPboundapproximate}, \text{ }\eqref{RelaxSCABapproximate2}, \text{ }\eqref{RelaxSCABapproximateall}, \text{ }\eqref{RelaxSCABCnewvariable1lowerbound1}, \text{ }\eqref{RelaxSCABCnewvariable1lowerbound2},
\end{align}
\end{subequations}
where ${\mathcal{B}} =
\left\{ {{{\bm{w}}_{{\rm{D}},k}}, {{\bm{u}}_{{\rm{U}},j,z}}, {P_{{\rm{U}},j}}, \tilde \mu _{\text{D},k}, {{{\tilde \rho}_{{\rm{D}},k}}}, {\bar {\tilde \rho} } _{{\rm{D}},k,l}, {\beta _{{\rm{U}},j,j'}},{{\tilde P}_{{\rm{D}},l}},} \right.\\
\quad\quad\quad\left. {{{\bar {\tilde P}}_{{\rm{D}},l,j}},{\chi _{{\rm{U}},j}},{\chi _{{\rm{D}},k}}} \right\}.$

The proposed SPCA algorithm for problem \eqref{MSRAS1} is outlined in Algorithm \ref{alg4}.


\begin{algorithm}[htb]
\caption{Solution to Problem \eqref{MSRAS1} with the Iterative SPCA-based Algorithm.}
\label{alg4}
\begin{algorithmic}[1] 
\REQUIRE ~~\\ {initialization:  $ \left\{{\mathcal{B}}\right\}^{( 0 )}$.
\\$\text{Set } {n} {\rm{ }} \to {\rm{ 0}} $}.
\REPEAT
\STATE Solve problem \eqref{RelaxA} to find optimal solutions $ \left\{{\mathcal{B}}\right\}^{\dag}$ with the current feasible point $ \left\{{\mathcal{B}}\right\}^{\left( n \right)}$.
\STATE Update $ \left\{{\mathcal{B}}\right\}^{{\left( n+1 \right)}} \to  \left\{{\mathcal{B}}\right\}^{{\dag}}$.
\STATE $\text{Set } {n} {\rm{ }} \to {\rm{n+1}} $.
\UNTIL Convergence.
\RETURN The optimal solutions $ \left\{ P^{\dag} _{\text{U},j}, \bm{w}^{\dag}_{\text{D},k}, {{{\bm{u}}^{\dag}_{\text{U},j,z}}}\right\}$.
\end{algorithmic}
\end{algorithm}

Although the proposed SPCA algorithm has a lower complexity compared to the SDR-BCD algorithm, the initialization
condition of the SPCA algorithm is more stringent. For the SDR-BCD algorithm, the received beamforming vector  $\bm{u}^{\left( {0} \right)}_{\text{U},j,z}$ can be initialized with an arbitrary vector.
Meanwhile, the SPCA algorithm must be initialized with a feasible point, and an infeasible initialization point will make the problem incapable of being solved. One can refer to \cite{boyd2004convex} to find a feasible solution.

\begin{proposition}
\label{Thm_2}
The proposed iterative  SPCA-based algorithm is guaranteed to monotonically converge. Moreover, the converged solution
generated by the proposed  SPCA-based algorithm is a KKT solution of problem \eqref{Relax}.
\begin{proof}
Please refer to Appendix B.
\end{proof}
\end{proposition}

\subsection{Stage \uppercase\expandafter{\romannumeral2}: Solution to Problem \eqref{MSRAS2} with the Iterative SPCA-based Algorithm}
Similar to Section \uppercase\expandafter{\romannumeral3}, given the user association,  problem \eqref{MSRAS2} can be reformulated as
\begin{subequations}\label{RelaxB}
\begin{align}
\mathop {\max }\limits_{\left\{\mathcal{B} \right\}}& \quad\quad{\vartheta ^{\left( 0 \right)}} \\ \label{RelaxA1}
s.t.\quad&\text{ }\eqref{MSR1}, \text{ } \eqref{MSR2}, \text{ }\eqref{MSR6}, \text{ } \eqref{Relax3}, \text{ } \eqref{Soc1},\text{ } \eqref{Soc2}, \\
&\text{ }\eqref{DRelaxR}, \text{ }\eqref{Qos}, \text{ }\eqref{DWMMSEE1},  \text{ }\eqref{DWMMSEE4}, \text{ }\eqref{DWMMSEF},\\
&\text{ }\eqref{FirstOrderCL}, \text{ }\eqref{FirstOrderCLResortrewrite},  \text{ }\eqref{URelaxUPboundapproximate}, \text{ }\eqref{RelaxSCABCnewvariable1lowerbound1}, \text{ }\eqref{RelaxSCABCnewvariable1lowerbound2},\\
&\sum\limits_{k \in {{\cal K}_{{\rm{D}},l}}}^{} {{{\tilde \rho}_{{\rm{D}},k}}}  \le {C_{{\rm{D}},l}}.
\end{align}
\end{subequations}

The iterative SPCA algorithm for problem \eqref{MSRAS2} is
summarized in Algorithm \ref{alg5}, where ${\mathcal{B}_0} =\left\{ {{\bm{w}}_{{\rm{D}},k}},{{\bm{u}}_{{\rm{U}},j,z}}, {P_{{\rm{U}},j}}, \tilde \mu _{\text{D},k},  \right.\\
\left. {{{\tilde \rho}_{{\rm{D}},k}}},{\beta _{{\rm{U}},j,j'}},
 {{{\tilde P}_{{\rm{D}},l}},{{\bar {\tilde P}}_{{\rm{D}},l,j}},{\chi _{{\rm{U}},j}},{\chi _{{\rm{D}},k}}} \right\}$.

The overall two-stage SPCA-based algorithm for problem \eqref{MSR} is summarized
in Algorithm \ref{alg6}.
\begin{algorithm}[htb]
\caption{Solution to Problem \eqref{MSRAS2} with the Iterative SPCA-based Algorithm.}
\label{alg5}
\begin{algorithmic}[1] 
\REQUIRE ~~\\ {Generate initial feasible points $ \left\{{\mathcal{B}_0}\right\}^{( 0 )}$.
\\$\text{Set } {n} {\rm{ }} \to {\rm{ 0}} $}.
\REPEAT
\STATE Solve problem \eqref{RelaxB} to find optimal solutions $ \left\{{\mathcal{B}_0}\right\}^{\dag }$.

\STATE Run 3 and 4 of Algorithm \ref{alg4}.
\UNTIL Convergence.
\RETURN The optimal solutions $ \left\{ P^{\dag} _{\text{U},j}, \bm{w}^{\dag}_{\text{D},k}, {{{\bm{u}}^{\dag}_{\text{U},j,z}}}\right\}$.
\end{algorithmic}
\end{algorithm}

\begin{algorithm}[htb]
\caption{ Overall Iterative SPCA-based Algorithm.}
\label{alg6}
\begin{algorithmic}[1] 
\STATE Run Algorithm \ref{alg4} to obtain the DU-RAU association.
\STATE Run Algorithm \ref{alg5} to obtain the sparse beamforming and power control strategy.
\end{algorithmic}
\end{algorithm}

It can be shown that the  problem \eqref{RelaxA} has $2MKL+J$ real variables and has a complexity order of $O((2MKL+J)^{2}(2J(2M+3J+L+1)+KL(2MLK+2M+5)+4K(ML+K+J-1)+7K+2MK^{2}+2L)+3(2^{(N-1)}+\sum\limits_{t = 1}^{N-2} 2^t))$ \cite{wang2014outage}.

\section{Numerical Results}
\begin{table}\label{Simulation}
\centering 
\caption{Simulation Parameters}
\begin{tabular}{|c|c|}\hline
Radius & 60 m \\ \hline
Power constraint for RAU/User  & 1 W/0.5 W \\ \hline
No. of T-RAUs/R-RAUs (DUs/UUs) & 10/10 (5/5) \\ \hline
Path loss  & 128.1+37.6log10(d) \\ \hline
Lognormal shadowing/Rayleigh fading& 8 dB/0 dB \\ \hline
Noise power (${\sigma^2_{\text{U},z}}=\sigma _{{\rm{D}},k}^2=\sigma^2$) & -70 dBm \\ \hline
\end{tabular}
\end{table}

In this section, some numerical examples are evaluated to show the performance of the proposed algorithms under different system settings.
We consider an L-DAS in a circular area with the detailed simulation parameters listed in Table \uppercase\expandafter{\romannumeral2}.
For simplicity, we set equal backhaul constraints. We model the residual interference power between T-RAU $l$  and R-RAU $z$ as ${{\delta _{{\rm{IRI}},l,z}}} ={\Delta _{{\text{IRI}},l,z}}\sigma^2=\Delta\sigma^2$,
where ${\Delta _{{\text{IRI}},l,z}}={\Delta}, \forall l,z$  represents the ratio of the channel estimation error.

\begin{figure}[htbh]
\begin{center}
{\includegraphics[scale=0.46]{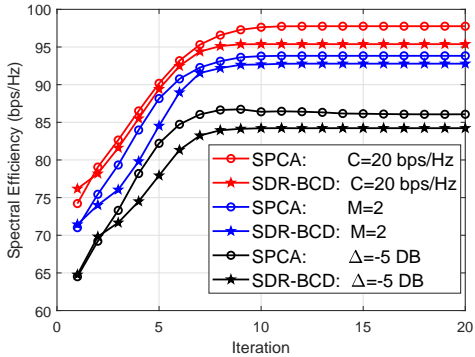}}
\end{center}
\caption{Convergence performance versus iteration.}
\label{Convergence}
\end{figure}

Fig. \ref{Convergence} shows the convergence behaviour of the proposed iterative SDR-BCD
and SPCA-based algorithms for problem \eqref{MSRAS1} when $C_{\text{D},l}=20$ \text{bps/Hz}, $M=2$ and $\Delta=-5$ \text{dB}, with the other fixed parameters. It can be observed that the proposed algorithms converge roughly within 10-15 iterations. Specifically, the SPCA-based algorithm achieves the best steady-state performance with monotonic convergence,
followed by the SDR-BCD-based algorithm.

\begin{figure}[htbh]
\begin{center}
{\includegraphics[scale=0.46]{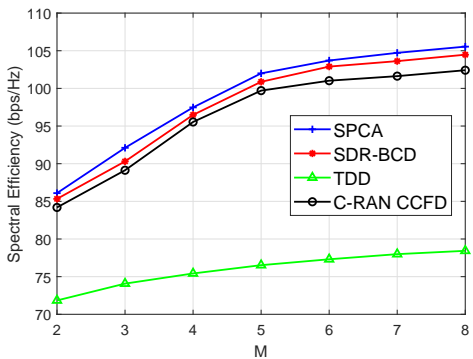}}
\end{center}
\caption{SE performance versus M.}
\label{MChange}
\end{figure}

Fig. \ref{MChange} illustrates the SE performance versus the number of RAU antennas $M$ for ${R_{{\text{D},\text{min}},k}}={R_{{\text{U},\text{min}},j}}=0.1$ \text{bps/Hz},  ${\Delta}=-5$ \text{dB}, and $C_{\text{D},l}=60$ \text{bps/Hz}. To clarify the problem, we compare our algorithms with the TDD and C-RAN CCFD schemes.
As expected, from Fig. \ref{MChange}, we can
find that the SE of all algorithms rapidly increases with the increase in the number of antennas.  Specifically, the SE of the SPCA- and SDR-BCD-based algorithms is higher than those of the TDD  and C-RAN FD schemes, approximately (22.87, 2.57) bps/Hz and (21.77, 1.48) bps/Hz, respectively.
It is also observed that the proposed algorithms achieve a close SE performance, and the SPCA algorithm achieves a better performance than the SDR-BCD  algorithm.

\begin{figure}[htbh]
\begin{center}
{\includegraphics[scale=0.46]{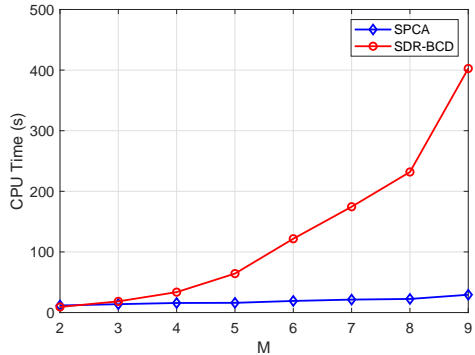}}
\end{center}
\caption{CPU time versus M.}
\label{CPUTime}
\end{figure}

Next, we compare the performance of the proposed algorithms in terms of CPU time (i.e., time complexity) over 100 problem instances. Fig. \ref{CPUTime} demonstrates the
CPU time consumption of the proposed algorithms versus the number of RAU transmit antennas M, with a fixed number of uplink users and downlink users and set ${R_{{\text{D},\text{min}},k}}={R_{{\text{U},\text{min}},j}}=0.1 \text{ bps/Hz}, {\Delta _{{\rm{IRI}},l,z}}=-10 \text{ dB},$ and $C_{\text{D},l}=60 \text{ bps/Hz}$. It can be seen from this figure that the time consumed by the SDR-BCD-based algorithm rapidly increases with M, while the time complexity of the SPCA-based algorithm merely
changes with different M. Moreover, the time consumed by both algorithms is close to each other when $M \le 4$, and the gap is obvious when $M > 4$. For example, when M=4, the time
consumed by the SDR-BCD-based algorithm is approximately 2.13 times that by the SPCA-based algorithm, while the increases are 4.01- and 13.72-fold when M=5 and M=9, respectively.

\begin{figure}[htbh]
\begin{center}
{\includegraphics[scale=0.46]{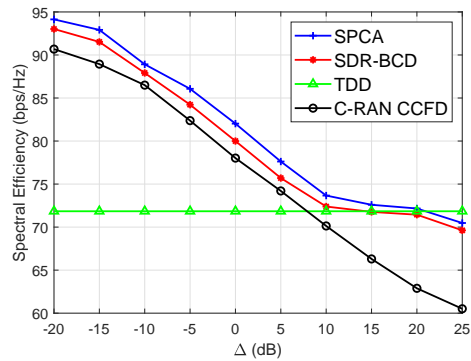}}
\end{center}
\caption{SE performance versus channel estimation error.}
\label{channeleatimation}
\end{figure}

Then, the SE performance versus channel estimation error for ${R_{{\text{D},\text{min}},k}}={R_{{\text{U},\text{min}},j}}=0.1$ \text{bps/Hz}, $M = 2$, and $C_{\text{D},l}=60$  \text{bps/Hz} is given in Fig. \ref{channeleatimation}. A general observation is that the proposed algorithms with the NAFD and C-RAN CCFD schemes can significantly improve the SE compared with the TDD scheme when ${{\delta _{{\rm{IRI}},l,z}}}$  is substantially suppressed, i.e., ${\Delta }$ is as small as possible.
Specifically, as shown in Fig. \ref{channeleatimation}, when ${\Delta } \le 15 $ \text{dB} and ${\Delta } \le 5$ \text{dB}, the proposed algorithms with NAFD and C-RAN CCFD perform better than that with TDD, respectively, and when ${\Delta}$ reduces to  $-20$ dB, the
total SE gains of the two schemes are $22.29 $ bps/Hz and $18.84$ bps/Hz higher than that of the TDD scheme, respectively. However, when ${\Delta _{{\rm{IRI}},l,z}}> 25 $ \text{dB} and ${\Delta _{{\rm{IRI}},l,z}}> 10 $ \text{dB}, the TDD system outperforms the NAFD and C-RAN CCFD systems.
In addition, we also find that the total SE gain of the SPCA-based algorithm at ${\Delta }=-20 $ \text{dB} is $20.47 $ \text{ bps/Hz} higher than that when ${\Delta _{{\rm{IRI}},l,z}}=10 $  \text{dB}, and the increase of the SE gain will be faster when ${\Delta } \le 10 $ \text{dB}.

 \begin{figure}[htbh]
\begin{center}
{\includegraphics[scale=0.46]{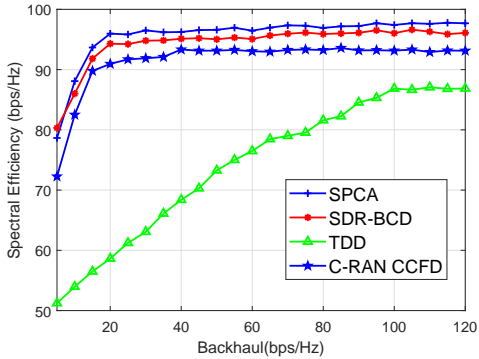}}
\end{center}
\caption{SE performance versus backhaul constraint for each RAU.}
\label{Backhaul}
\end{figure}
Finally, in Fig. \ref{Backhaul}, we illustrate the SE performance versus backhaul constraint for each RAU when ${R_{{\text{D},\text{min}},k}}={R_{{\text{U},\text{min}},j}}=0.1$ \text{bps/Hz}, $M = 4$, and ${\Delta _{{\rm{{\rm{IRI}}}},l,z}}= -10 $ \text{dB}.
 As expected, from Fig. \ref{Backhaul}, we can find that all algorithms that include the TDD and C-RAN CCFD schemes achieve higher SE performance as the backhaul constraint increases. However, our proposed algorithms obtain a higher SE gain than the TDD scheme, and compared to the TDD scheme, the proposed algorithms and the C-RAN CCFD scheme can easily achieve stable SE gain.

  For example, the SE of the SPCA-based and SDR-BCD-based algorithms reaches 95.97 bps/Hz - 97.69 bps/Hz and  94.30 bps/Hz - 96.10 bps/Hz,  while that of the TDD scheme reaches 58.64 bps/Hz - 86.87 bps/Hz, when the backhaul increases from 20 bps/Hz to 120 bps/Hz. The reason can be explained as follows. When the backhaul constraint is 20 bps/Hz, the downlink transmission is suppressed due to the  downlink backhaul constraint, while the uplink can achieve a higher SE gain given the lower interferences caused by T-RAUs.  However, in the TDD scheme, the downlink and uplink transmission operate in  different T-F slots, and the low level of the downlink backhaul constraint will significantly suppress the downlink transmission while the uplink transmission operates in another T-F slot with a regular level.
   Furthermore, when the backhaul constraint is lower than 20 bps/Hz, the SE gain of all the schemes are all at a low level because most of the backhaul links operate
  in an unsatisfactory status or series interferences occur between downlink and uplink transmission.


\section{Conclusion}
In this paper, we have studied the joint sparse beamforming and power control for an L-DAS with NAFD under finite backhaul and QoS constraints.
To deal with the finite backhaul, we have designed downlink sparse beamforming by approximating the $l_0$
norm as a concave function and proposed a two-stage iterative algorithm.
Two approaches were proposed to handle the highly non-convex joint transceiver design problem with a performance and complexity trade-off.
In the first approach, we proposed an iterative SDR-BCD algorithm with alternatively fixed transmitter and receiver for
distributed design of the optimization variables.
 In the second design approach, we proposed an iterative SPCA
algorithm to jointly deal with the non-convex constraints by approximating them as convex ones.
 The simulation results showed that the proposed
algorithms have a superior performance over the TDD scheme when the IRI is suppressed to a low stage.

\ifCLASSOPTIONcaptionsoff
  \newpage
\fi
\appendices
\section{Proof of proposition \ref{Thm_1}}
Let us first start with the convergence proof of Algorithm \ref{alg1} when $\bm{u}_{\rm{U}}$ are fixed.
Thus, we first define the two important properties as
\begin{equation}\label{tight}
\begin{aligned}
{h^{\left( {n + 1} \right)}}\left( {{{\bm{Q}}^{\left( n \right)}},{{\bm{P}}^{\left( n \right)}}} \right){\rm{ = }}h\left( {{{\bm{Q}}^{\left( n \right)}},{{\bm{P}}^{\left( n \right)}}} \right),
\end{aligned}
\end{equation}
\begin{equation}\label{tight2}
\begin{aligned}
\nabla {h^{\left( {n + 1} \right)}}\left( {{{\bm{Q}}^{\left( n \right)}},{{\bm{P}}^{\left( n \right)}}} \right){\rm{ = }}\nabla h\left( {{{\bm{Q}}^{\left( n \right)}},{{\bm{P}}^{\left( n \right)}}} \right),
\end{aligned}
\end{equation}
where property \eqref{tight} means that the optimal value of \eqref{MSRQfh} at the $(n+1)$th iteration is tight to iteration $n$ when  $\left( {{\bm{Q}},{\bm{P}}} \right) = \left( {{{\bm{Q}}^{\left( n \right)}},{{\bm{P}}^{\left( n \right)}}} \right)$.

Let ${\omega ^{\left( {n + 1} \right)}}$ be the optimal objective of \eqref{MSRQfh} at $(n+1)$th iteration; based on \cite{nguyen2014spectral}, we obtain
\begin{equation}\label{Cov}
\begin{aligned}
{\omega ^{\left( {n + 1} \right)}}& = f\left( {{{\bm{Q}}^{\left( {n + 1} \right)}},{{\bm{P}}^{\left( {n + 1} \right)}}} \right) - {h^{\left( {n + 1} \right)}}\left( {{{\bm{Q}}^{\left( {n + 1} \right)}},{{\bm{P}}^{\left( {n + 1} \right)}}} \right)  \\
 &\ge f\left( {{{\bm{Q}}^{\left( n \right)}},{{\bm{P}}^{\left( n \right)}}} \right) - {h^{\left( {n - 1} \right)}}\left( {{{\bm{Q}}^{\left( n \right)}},{{\bm{P}}^{\left( n \right)}}} \right) = {\omega ^{\left( n \right)}}.
\end{aligned}
\end{equation}

Inequation \eqref{Cov} shows that the sequence ${\omega ^{\left( {n + 1} \right)}}$ is nondecreasing. Moreover, ${\omega ^{\left( {n + 1} \right)}}$ is bounded above due to the
limited transmit power; thus, it is guaranteed to converge.

Then, let $\zeta$  denote the feasible set of \eqref{MSRQfhn}.
Constraints \eqref{MSRQ1}, \eqref{MSRQ2}, \eqref{MSRQ3}, \eqref{MSRQ5} and \eqref{MSRQ6} are all of convex form due to the  SDR approximation.
The solution satisfies the convex constraint in problem \eqref{MSRQfhn} and must satisfy the corresponding constraints in problem \eqref{MSRQfh}.
Furthermore, we have used the upper bound to approximate non-convex constraint  \eqref{MSRQ5}, as shown in \eqref{MSRQfhnS},\text{ } \eqref{BackhaulBCDRate1}  and \eqref{URelaxSCAR};
thus, any feasible solution to problem \eqref{MSRQfhn} satisfies all the constraints of problem \eqref{MSRQfh}.

The
alternative optimization with MMSE receiver $\bm{u}_U$ can achieve the same optimal value for problem \eqref{MSRQfh} and converges to a stationary point of \eqref{MSRQfh} \cite{shen2018fractional}; hence, the alternative optimization with
 MMSE receiver $\bm{u}_{\rm{U}}$ will not affect the convergence of problem \eqref{MSRQfh}, which thus completes the proof.

\section{Proof of proposition \ref{Thm_2}}
The convergence proof of Algorithm \ref{alg4} follows the same spirit as the proof of proposition \ref{Thm_1}. In the following, we try to prove that the converged solution satisfies the KKT conditions of problem \eqref{ReForm}.

Define a non-convex constraint ${f_{nc}}\left( \bm b \right) \le 0$, which is iteratively approximated by the convex constraint ${f_c}\left( {\bm b,\bm \tilde  b} \right) \le 0$, where $\bm \tilde  b$ is the  optimal solution to the approximated problem in the previous iteration.
Additionally, we suppose the two problems satisfy the following conditions:

C1: ${f_c}\left( {\bm{b},\bm{\tilde b}} \right) \ge {f_{nc}}\left( \bm b \right)$.

C2: ${f_c}\left( {\bm {\tilde b},\bm {\tilde b}} \right) = {f_{nc}}\left( {\bm {\tilde b}} \right)$.

C3: ${\nabla _{\bm b = \bm {\tilde b}}}{f_c}\left( {\bm b, \bm {\tilde b}} \right) = {\nabla _{\bm b = \bm {\tilde b}}}{f_{nc}}\left(\bm b \right)$.

C4: The approximated problem satisfies Slater's \text{ }\text{ }\text{ }\text{ }\text{ }\text{ }\text{ }\text{ }condition.

Then, the  successive convex approximation algorithm  can always yield a solution satisfying the KKT conditions of the problem \cite{marks1978general,nguyen2014spectral}.
Given that in Algorithm \ref{alg4}, we use the SPCA method result and the first-order Taylor expansion as the upper bound or lower bound to approximate the non-convex constraints
in problem \eqref{ReForm}, it is easy to check that the series of constraints
\eqref{Relax3}, \eqref{Soc1}, \eqref{Soc2}, \eqref{DRelaxR}, \eqref{Qos}, \eqref{DWMMSEE1}, \eqref{DWMMSEE4}, \eqref{DWMMSEF}, \eqref{FirstOrderCL}, \eqref{FirstOrderCLResortrewrite}, \eqref{URelaxUPboundapproximate}, \eqref{RelaxSCABapproximate2}, \eqref{RelaxSCABapproximateall} and \eqref{RelaxSCABCnewvariable1lowerbound} are
approximations to constraints \eqref{ReFor4}-\eqref{ReFor6} satisfying the conditions given in C1-C3.

Finally, we will show that problem \eqref{RelaxA} satisfies Slater's condition, i.e., condition C4. Specifically, let $\left\{ {\mathcal{B}} \right\}^{\left( n \right)}$
denote a feasible solution to problem \eqref{RelaxA} in the $n$-th iteration.
Given any sequences $\left\{ {{\varsigma _{\rm{D},1}}, \cdots ,{\varsigma _{{\rm{D}},K}}} \right\}$ and $\left\{ {{\varsigma _{\rm{U},1}}, \cdots ,{\varsigma _{{\rm{U}},J}}} \right\}$ that satisfy
$ {\varsigma _{{\rm{D}},{{k}}}} \le 1, \forall k$ and $ {\varsigma _{{\rm{U}},j}} \le 1, \forall j$, respectively,
then we consider the solution as 
 \begin{equation}\label{Solution}
\begin{aligned}
\left\{ {{{\bm{w}}_{{\rm{D}},k}}{\rm{ = }}{\varsigma _{{\rm{D}},{{k}}}}{\bm{w}}_{{\rm{D}},k}^{\left( n \right)},{P_{{\rm{U}},j}} = {\varsigma _{{\rm{U}},j}}P_{{\rm{U}},j}^{\left( n \right)},}
 {\left\{ {{{\cal B}_1}} \right\}{\rm{ = }}{{\left\{ {{{\cal B}_1}} \right\}}^{\left( n \right)}}} \right\},
\end{aligned}
\end{equation}
where
${\mathcal{B}}_1 =\left\{  {{\bm{u}}_{{\rm{U}},j,z}},  \tilde \mu _{\text{D},k}, {{{\tilde \rho}_{{\rm{D}},k}}}, {\bar {\tilde \rho} } _{{\rm{D}},k,l},{\beta _{{\rm{U}},j,j'}},{{\tilde P}_{{\rm{D}},l}}, {{\bar {\tilde P}}_{{\rm{D}},l,j}}, \right. \\
\left.{\chi _{{\rm{U}},j}},{\chi _{{\rm{D}},k}} \right\}$.

With this solution, we can confirm \eqref{MSR2}, \eqref{MSR3}, \eqref{DWMMSEE1}, \eqref{DWMMSEE4} and  \eqref{DWMMSEF} as
\begin{equation}\label{Slate}
\begin{aligned}
\sum\limits_{k \in {{\cal K}}} {{{\left\| {{{\bm{w}}_{{\rm{D}},l,k}}} \right\|}^2}}  = \sum\limits_{k \in {{\cal K}}} {\varsigma _{{\rm{D}},{\rm{k}}}^2{{\left\| {{\bm{w}}_{{\rm{D}},l,k}^{\left( n \right)}} \right\|}^2}}  \le {{{\rm{\bar P}}}_{{\rm{D}},l}},\forall l,
\end{aligned}
\end{equation}
\begin{equation}\label{Slate1}
\begin{aligned}
{P_{{\rm{U}},j}} = {\varsigma _{{\rm{U}},j}}P_{{\rm{U}},j}^{\left( n \right)} \le P_{{\rm{U}},j}^{\left( n \right)} \le {{\bar P}_{{\rm{U}},j}},\forall j,
\end{aligned}
\end{equation}
\begin{small}
\begin{equation}\label{Slate3}
\begin{aligned}
\sum\limits_{k \in \mathcal{K}} {{{\left\| {{{\bm{w}}_{{\rm{D}},l,k}}} \right\|}^2}}  = \sum\limits_{k \in \mathcal{K}}{\varsigma _{{\rm{D}},l,k}^2{{\left\| {{\bm{w}}_{{\rm{D}},l,k}^{\left( n \right)}} \right\|}^2}}  < \sum\limits_{k \in \mathcal{K}} {{{\left\| {{\bm{w}}_{{\rm{D}},l,k}^{\left( n \right)}} \right\|}^2}}  \le \tilde P_{{\rm{D}},l}^{\left( n \right)},
\end{aligned}
\end{equation}
\end{small}
\begin{equation}\label{Slate4}
\begin{aligned}
{{\tilde P}_{{\rm{D}},l}}\delta _{{\rm{IRI}},l,z}^2{\left\| {{{\bm{u}}_{{\rm{U}},j,z}}} \right\|^2} = \tilde P_{{\rm{D}},l}^{\left( n \right)}\delta _{{\rm{IRI}},l,z}^2{\left\| {{\bm{u}}_{{\rm{U}},j,z}^{\left( n \right)}} \right\|^2} \le {\left( {\bar {\tilde P}_{{\rm{D}},l,j}^2} \right)^{\left( n \right)}},
\end{aligned}
\end{equation}
\begin{equation}\label{Slate5}
\begin{aligned}
{{\tilde P}_{{\rm{D}},l}} = \tilde P_{{\rm{D}},l}^{\left( n \right)} \le {{{\rm{\bar P}}}_{{\rm{D}},l}}.
\end{aligned}
\end{equation}

In the same spirit, we can confirm that the solution given in  \eqref{Solution} is a strictly feasible solution to
problem \eqref{RelaxA}. Therefore, Slater's condition holds for problem \eqref{RelaxA} \cite{boyd2004convex}.
Therefore, the solution obtained by Algorithm \ref{alg3} satisfies the KKT condition
of problem \eqref{ReForm}.

\bibliographystyle{IEEEtran}
\bibliography{Joint_Sparse_Beamforming_and_Power_Control_for_Large-scale_DAS_with_Network_Assistant_Full_Duplex}

\begin{thebibliography}{10}
\providecommand{\url}[1]{#1}
\csname url@samestyle\endcsname
\providecommand{\newblock}{\relax}
\providecommand{\bibinfo}[2]{#2}
\providecommand{\BIBentrySTDinterwordspacing}{\spaceskip=0pt\relax}
\providecommand{\BIBentryALTinterwordstretchfactor}{4}
\providecommand{\BIBentryALTinterwordspacing}{\spaceskip=\fontdimen2\font plus
\BIBentryALTinterwordstretchfactor\fontdimen3\font minus
  \fontdimen4\font\relax}
\providecommand{\BIBforeignlanguage}[2]{{%
\expandafter\ifx\csname l@#1\endcsname\relax
\typeout{** WARNING: IEEEtran.bst: No hyphenation pattern has been}%
\typeout{** loaded for the language `#1'. Using the pattern for}%
\typeout{** the default language instead.}%
\else
\language=\csname l@#1\endcsname
\fi
#2}}
\providecommand{\BIBdecl}{\relax}
\BIBdecl

\bibitem{sun2015d2d}
H.~Sun, M.~Wildemeersch, M.~Sheng, and T.~Q. Quek, ``{D2D} enhanced
  heterogeneous cellular networks with dynamic {TDD},'' \emph{IEEE Transactions
  on Wireless Communications}, vol.~14, no.~8, pp. 4204--4218, 2015.

\bibitem{sabharwal2014band}
A.~Sabharwal, P.~Schniter, D.~Guo, D.~W. Bliss, S.~Rangarajan, and R.~Wichman,
  ``In-band full-duplex wireless: Challenges and opportunities.'' \emph{IEEE
  Journal on Selected Areas in Communications}, vol.~32, no.~9, pp. 1637--1652,
  2014.

\bibitem{nguyen2014spectral}
D.~Nguyen, L.-N. Tran, P.~Pirinen, and M.~Latva-aho, ``On the spectral
  efficiency of full-duplex small cell wireless systems,'' \emph{IEEE
  Transactions on Wireless Communications}, vol.~13, no.~9, pp. 4896--4910,
  2014.

\bibitem{li2017spectral}
Y.~Li, P.~Fan, A.~Leukhin, and L.~Liu, ``On the spectral and energy efficiency
  of full-duplex small-cell wireless systems with massive {MIMO},'' \emph{IEEE
  Transactions on Vehicular Technology}, vol.~66, no.~3, pp. 2339--2353, 2017.

\bibitem{bi2018fractional}
W.~Bi, L.~Xiao, X.~Su, and S.~Zhou, ``Fractional full duplex cellular network:
  a stochastic geometry approach,'' \emph{Science China Information Sciences},
  vol.~61, no.~2, p. 022302, 2018.

\bibitem{song2017antenna}
K.~Song, C.~Li, Y.~Huang, and L.~Yang, ``Antenna selection for two-way full
  duplex massive {MIMO} networks with amplify-and-forward relay,''
  \emph{Science China Information Sciences}, vol.~60, no.~2, p. 022308, 2017.

\bibitem{thomsen2016compflex}
H.~Thomsen, P.~Popovski, E.~De~Carvalho, N.~K. Pratas, D.~M. Kim, and
  F.~Boccardi, ``{CoMPflex}: {CoMP} for in-band wireless full duplex,''
  \emph{IEEE Wireless Communications Letters}, vol.~5, no.~2, pp. 144--147,
  2016.

\bibitem{xin2017antenna}
Y.~Xin, R.~Zhang, D.~Wang, J.~Li, L.~Yang, and X.~You, ``Antenna clustering for
  bidirectional dynamic network with large-scale distributed antenna systems,''
  \emph{IEEE Access}, vol.~5, pp. 4037--4047, 2017.

\bibitem{Dongming2019}
\BIBentryALTinterwordspacing
D.~Wang, M.~Wang, P.~Zhu, J.~Li, J.~Wang, and X.~You, ``Performance of
  network-assisted full-duplex for cell-free massive {MIMO},'' \emph{Submitted
  to IEEE Trans. on Communications}, May 2019. [Online]. Available:
  \url{https://arxiv.org/abs/1905.11107}
\BIBentrySTDinterwordspacing

\bibitem{dai2014sparse}
B.~Dai and W.~Yu, ``Sparse beamforming and user-centric clustering for downlink
  cloud radio access network,'' \emph{IEEE Access}, vol.~2, pp. 1326--1339,
  2014.

\bibitem{dai2016energy}
------, ``Energy efficiency of downlink transmission strategies for cloud radio
  access networks,'' \emph{IEEE Journal on Selected Areas in Communications},
  vol.~34, no.~4, pp. 1037--1050, 2016.

\bibitem{tabassum2016analysis}
H.~Tabassum, A.~H. Sakr, and E.~Hossain, ``Analysis of massive {MIMO}-enabled
  downlink wireless backhauling for full-duplex small cells,'' \emph{IEEE
  Transactions on Communications}, vol.~64, no.~6, pp. 2354--2369, 2016.

\bibitem{dong2015energy}
Y.~Dong, H.~Zhang, M.~J. Hossain, J.~Cheng, and V.~C. Leung, ``Energy efficient
  resource allocation for {OFDMA} full duplex distributed antenna systems with
  energy recycling,'' in \emph{2015 IEEE Global Communications Conference
  (GLOBECOM)}.\hskip 1em plus 0.5em minus 0.4em\relax IEEE, 2015, pp. 1--6.

\bibitem{cirik2018fronthaul}
A.~C. Cirik, O.~Taghizadeh, L.~Lampe, and R.~Mathar, ``Fronthaul compression
  and precoding design for {MIMO} full-duplex cognitive radio networks,'' in
  \emph{2018 IEEE Wireless Communications and Networking Conference
  (WCNC)}.\hskip 1em plus 0.5em minus 0.4em\relax IEEE, 2018, pp. 1--6.

\bibitem{li2018spectral}
X.~Li, C.~He, and J.~Zhang, ``Spectral efficiency and energy efficiency of
  bidirectional distributed antenna systems with user centric virtual cells,''
  \emph{IEEE Access}, vol.~6, pp. 49\,886--49\,895, 2018.

\bibitem{chen2016green}
L.~Chen, F.~R. Yu, H.~Ji, B.~Rong, X.~Li, and V.~C. Leung, ``Green full-duplex
  self-backhaul and energy harvesting small cell networks with massive
  {MIMO},'' \emph{IEEE Journal on Selected Areas in Communications}, vol.~34,
  no.~12, pp. 3709--3724, 2016.

\bibitem{jiang2017max}
Y.~Jiang, F.~C. Lau, I.~W.-H. Ho, H.~Chen, and Y.~Huang, ``{Max-Min} weighted
  downlink {SINR} with uplink {SINR} constraints for full-duplex {MIMO}
  systems,'' \emph{IEEE Transactions on Signal Processing}, vol.~65, no.~12,
  pp. 3277--3292, 2017.

\bibitem{tan2018virtual}
Z.~Tan, F.~R. Yu, X.~Li, H.~Ji, and V.~C. Leung, ``Virtual resource allocation
  for heterogeneous services in full duplex-enabled {SCNs} with mobile edge
  computing and caching,'' \emph{IEEE Transactions on Vehicular Technology},
  vol.~67, no.~2, pp. 1794--1808, 2018.

\bibitem{liu2017cross}
L.~Liu and W.~Yu, ``Cross-layer design for downlink multihop cloud radio access
  networks with network coding.'' \emph{IEEE Trans. Signal Processing},
  vol.~65, no.~7, pp. 1728--1740, 2017.

\bibitem{kha2012fast}
H.~H. Kha, H.~D. Tuan, and H.~H. Nguyen, ``Fast global optimal power allocation
  in wireless networks by local {DC} programming,'' \emph{IEEE Transactions on
  Wireless Communications}, vol.~11, no.~2, pp. 510--515, 2012.

\bibitem{tran2012fast}
L.-N. Tran, M.~F. Hanif, A.~T{\"o}lli, and M.~Juntti, ``Fast converging
  algorithm for weighted sum rate maximization in multicell {MISO} downlink,''
  \emph{IEEE Signal Process. Lett.}, vol.~19, no.~12, pp. 872--875, 2012.

\bibitem{beck2010sequential}
A.~Beck, A.~Ben-Tal, and L.~Tetruashvili, ``A sequential parametric convex
  approximation method with applications to nonconvex truss topology design
  problems,'' \emph{Journal of Global Optimization}, vol.~47, no.~1, pp.
  29--51, 2010.

\bibitem{algorithms1997theory}
Y.~Ye, ``Theory and analysis,'' 1997.

\bibitem{lobo1998applications}
M.~S. Lobo, L.~Vandenberghe, S.~Boyd, and H.~Lebret, ``Applications of
  second-order cone programming,'' \emph{Linear algebra and its applications},
  vol. 284, no. 1-3, pp. 193--228, 1998.

\bibitem{boyd2004convex}
S.~Boyd and L.~Vandenberghe, \emph{Convex optimization}.\hskip 1em plus 0.5em
  minus 0.4em\relax Cambridge university press, 2004.

\bibitem{zhou2018energy}
X.~Zhou and Q.~Li, ``Energy efficiency for {SWIPT in MIMO} two-way
  amplify-and-forward relay networks,'' \emph{IEEE Transactions on Vehicular
  Technology}, 2018.

\bibitem{wang2014outage}
K.-Y. Wang, A.~M.-C. So, T.-H. Chang, W.-K. Ma, and C.-Y. Chi, ``Outage
  constrained robust transmit optimization for multiuser {MISO} downlinks:
  Tractable approximations by conic optimization,'' \emph{IEEE Transactions on
  Signal Processing}, vol.~62, no.~21, pp. 5690--5705, 2014.

\bibitem{shen2018fractional}
K.~Shen and W.~Yu, ``Fractional programming for communication systems¡ªpart
  {II}: Uplink scheduling via matching,'' \emph{IEEE Transactions on Signal
  Processing}, vol.~66, no.~10, pp. 2631--2644, 2018.

\bibitem{marks1978general}
B.~R. Marks and G.~P. Wright, ``A general inner approximation algorithm for
  nonconvex mathematical programs,'' \emph{Operations research}, vol.~26,
  no.~4, pp. 681--683, 1978.

\end{thebibliography}

\end{document}